\documentclass[reprint]{JASA}
\usepackage{epstopdf}
\usepackage{multirow}
\usepackage{setspace}

\DeclareMathOperator\erf{erf}
\newcommand*\patchAmsMathEnvironmentForLineno[1]{%
  \expandafter\let\csname old#1\expandafter\endcsname\csname #1\endcsname
  \expandafter\let\csname oldend#1\expandafter\endcsname\csname end#1\endcsname
  \renewenvironment{#1}%
     {\linenomath\csname old#1\endcsname}%
     {\csname oldend#1\endcsname\endlinenomath}}%
\newcommand*\patchBothAmsMathEnvironmentsForLineno[1]{%
  \patchAmsMathEnvironmentForLineno{#1}%
  \patchAmsMathEnvironmentForLineno{#1*}}%
\AtBeginDocument{%
\patchBothAmsMathEnvironmentsForLineno{equation}%
\patchBothAmsMathEnvironmentsForLineno{align}%
\patchBothAmsMathEnvironmentsForLineno{flalign}%
\patchBothAmsMathEnvironmentsForLineno{alignat}%
\patchBothAmsMathEnvironmentsForLineno{gather}%
\patchBothAmsMathEnvironmentsForLineno{multline}%
}
\begin{document}
\title[Resolution Dependence of Scattered Field]{Resolution dependence of rough surface scattering using a power law roughness spectrum}
\thanks{Published in the Journal of the Acoustical Society of America. The published version of this preprint can be found at \url{https://doi.org/10.1121/10.0002974}}
\author{Derek R. Olson}
\email{dolson@nps.edu}
\affiliation{Oceanography Department, Naval Postgraduate School, Monterey, CA 93943}
\author{Anthony P. Lyons}
\affiliation{University of New Hampshire, Durham, NH 03824}

%
%
%
\begin{abstract}
Contemporary high-resolution sonar systems use broadband pulses and long arrays to achieve high resolution. It is important to understand effects that high-resolution sonar systems might have on quantitative measures of the scattered field due to the seafloor. A quantity called the broadband scattering cross section is defined, appropriate for high-resolution measurements. The dependence of the broadband scattering cross section, $\sigma_{bb}$ and the scintillation index, $SI$ on resolution was investigated for one-dimensional rough surfaces with power-law spectra and backscattering geometries. Using integral equations and Fourier synthesis, no resolution dependence of $\sigma_{bb}$ was found. The incoherently-averaged frequency-domain scattering cross section has negligible bandwidth dependence. $SI$ increases as resolution increases, grazing angle decreases, and spectral strength increases. This trend is confirmed for center frequencies of 100 kHz and 10 kHz, as well as for power-law spectral exponents of 1.5, 2, and 2.5. The hypothesis that local tilting at the scale of the acoustic resolution is responsible for intensity fluctuations was examined using a representative model for the effect of slopes (inspired by the composite roughness approximation). It was found that slopes are responsible in part for the fluctuations, but other effects, such as multiple scattering and shadowing may also play a role.
\end{abstract}
\maketitle

\section{Introduction}
\label{sec:introduction}
Theoretical treatment of wave scattering from rough interfaces is generally performed using an incident monochromiatic plane wave, which has a single direction and exists over infinite spatial extent. However, experimental measurements of the scattered field often employ broadband pulses to achieve high spatial resolution - desirable for seafloor mapping or target detection. Performance of such systems typically depends on the mean intensity of the scattered field from the seafloor, and more generally its probability density function. For scattering from one-dimensional (1D) roughness, the mean intensity is usually characterized in terms of the scattering cross-section per unit length per unit angle\footnote{or per unit area per unit solid angle for two-dimensional (2D) rough interfaces}, $\sigma$  -- hereafter referred to as the ``cross section'', ``scattering cross section,'' or ``scattering strength'' for the decibel version. Variability of the scattered intensity is often characterized using the scintillation index, $SI$ \cite{tatarski,ishimaru_volII}.

The scattering cross section for a frequency-domain incident field for one-dimensional roughness is defined as \cite{thorsos_1988}
\begin{align}
	\sigma_f = \frac{\langle I_s \rangle R}{I_i L_{eff}}
	\label{eq:planeWaveSigmaDefinition}
\end{align}
where $\langle I_s \rangle= \langle |p_s(f)|^2\rangle /(2\rho_0 c)$ is the mean scattered intensity in the far-field, $I_i=|p_0|^2/(2\rho_0 c)$ is the incident intensity in the direction of the incident wave vector, $\rho_0$ is the ambient density, $p_0$ is the complex amplitude of the incident plane wave, $L_{eff}$ is the effective ensonified length of the incident field, and $R$ is the distance between a patch of rough interface and the receiver location. Note that this definition is valid only for geometries with well-defined incident and scattered field directions and for a single frequency. Strictly, this definition of the scattering cross section is only true in the limit as the ensonified length becomes large compared to all length scales of interest (i.e. outer scale of the rough surface, or the acoustic wavelength), since monochromatic plane waves interact with the entire rough surface. This definition is often used for narrowband\footnote{In this work, when the bandwidth of the signal is a tenth of the center frequency or smaller, it is considered to be narrowband.} incident fields that provide a good approximation to a single-frequency tone.

In this work, a quantity termed the ``broadband scattering cross section'' is investigated, that is appropriate for cases with short pulses\footnote{To be consistent with the criterion for narrowband defined above, a ``short'' pulse is defined to contain ten or fewer cycles.}. For a plane-wave rectangular pulse of length $\tau$, this is defined as
\begin{align}
	\sigma_{bb} = \frac{R}{c\tau/(\cos\theta_i + \cos\theta_s)} \frac{\langle I_s \rangle}{I_i}
\end{align}
where $c$ is the sound speed, $\theta_i$ is the incident grazing angle, and $\theta_s$ is the scattered grazing angle. This quantity is discussed more fully and derived in Sec.~\ref{sec:estimation}. In this work, the frequency-domain version of the scattering cross section is referred to as the ``cross section'', and the broadband version is $\sigma_{bb}$. The broadband scattering cross section may exhibit pulse-length dependence if the properties of the ensemble of rough surfaces vary with resolution, especially for high-resolution systems. This dependence is not possible for the frequency-domain cross section.


The interface scattering cross-section (or its broadband version) characterizes the mean scattered power from an interface, but a more general property of the scattered field is the probability density function (pdf) of the modulus of the complex pressure, termed the envelope pdf. The envelope pdf is connected to performance of target detection systems, and has potential utility for use in remote sensing of the environment using high resolution systems \cite{lyons_etal_2009,lyons_etal_2016,olson_etal_2019}. The statistical distribution of pressures resulting from scattering from a rough, homogeneous interface with Gaussian height statistics has commonly been assumed to be Gaussian for the real and imaginary components, and a Rayleigh distribution for the envelope \cite{jakeman_1980}). In this situation, the scintillation index, or normalized intensity variance is unity. For heavy-tailed statistics (with more frequent large amplitude events), the scintillation index is greater than unity.

Arguments for Rayleigh-distributed scattered pressure magnitude follow from the assumption of a large number of independent surface elements contributing to the scattered field \cite{jakeman_1980,abraham_lyons_2002}. So long as the ensonfied area of a Gaussian, homogeneous rough interface is large (so there are many independent scatterers contributing to the field), this assumption holds true. Another argument for Rayleigh magnitudes follows from perturbation theory and the interpretation in terms of Bragg scattering. In this framework, the scattered pressure is proportional to the amplitude spectrum of the roughness evaluated at the Bragg wavenumber, $2 k_w \cos(\theta_i)$, where $k_w$ is the acoustic wavenumber in the water column, and $\theta_i$ is the incident grazing angle. If the surface has Gaussian statistics in the spatial domain, then the wavenumber components will have Rayleigh-distributed magnitude via the central limit theorem. Therefore, the acoustic envelope pdf will be Rayleigh-distributed as well.

Contemporary high-resolution seafloor imaging systems, such as synthetic aperture sonar (SAS) have spatial resolutions on the order of the center wavelength \cite{fossum_etal_2008,bellettini_pinto_2009,pinto_2011,dillon_2018,sternlicht_etal_2016}. Small resolution cell sizes may result in ensembles that vary with the resolved area of the seafloor, thereby causing a departure from Rayleigh statistics. The resolution dependence of the scintillation index has implications for target detection performance, synthetic aperture autofocus algorithms (e.g. \cite{marston_plotnick_2015}), and preprocessing algorithms for automated target recognition \cite{Kwon2005,williams_2015,Galusha2018}.

It was observed in \cite{lyons_etal_2016} that measurements of the scintillation index from SAS images of homogeneous random rough interfaces had a strong dependence on range, which was interpreted as a result of modulation of the local slope by roughness components at the scale of the acoustic resolution or larger, which is called the local-tilting hypothesis. This interpretation uses a model for local slope modulation inspired by the composite roughness approximation \cite{mcdaniel_gorman_1983}. Combined with interpretations in \cite{lyons_etal_2016}, this effect results in a dependence of the scintillation index on the acoustic resolution, the underlying pixel statistics, range (through grazing angle), as well as roughness spectrum parameters. These interpretations, while plausible, suffer from a lack of experimental confirmation. In the electromagnetics literature, slope modulation was postulated to cause non-Rayleigh scattering by \cite{Valenzuela1971}, and \cite{Li2017}. In the specular direction, the Fresnel zone (a form of resolved area) has been shown to affect the scintillation index \cite{Yang1991}, although this situation contains a coherent component and is not germane to the current problem.

In this work, the question is examined of whether there is a dependence on resolution of the broadband scattering cross section and scintillation index. The acoustic resolution, $\Delta X$ is defined as the full width half maximum spatial extent of the square of the incident pulse envelope. For broadband signals, the temporal resolution is set by $1/(2aB_{3dB})$, where $B_{3dB}$ is the 3 dB full width bandwidth of the transmitted pulse\footnote{If phase coding and pulse compression are employed, this analysis deals with the resolution of the compressed, or matched-filtered pulse.}, and $a$ is a constant which depends on the shape of the pulse used. The spatial resolution is $c/(2aB_{3dB})$ for small grazing angles.

These questions were investigated through numerical solution of the Helmholtz-Kirchhoff integral equation for the scattered pressure using the boundary element method (BEM) \cite{sauter_schwab,wu_bem} using pressure-release boundary conditions. This method is similar to that used by \cite{thorsos_1988}. Fourier synthesis was used to construct the broadband scattered pressure at various spatial resolutions, and metrics were computed based on the scattered time-domain pressure. The numerical method detailed here can, in general, treat bistatic geometries, but only the monostatic case was examined. Comparisons were made to the ensemble averaged cross section performed in the frequency domain (i.e. computed at a single frequency, which is a good approximation of the monochromatic plane wave case). These simulations were performed for center frequencies of both 100 kHz and 10 kHz, and for one-dimensional rough surfaces with power law spectra, whose parameters are the spectral strength and spectral exponent.

Comparisons between numerical simulations and field experiments would require two-dimensional (2D) roughness with a three-dimensional (3D) scattering geometry. However, the numerical method has very high computational and memory requirements that made 3D simulations impossible at this time, and therefore a 2D geometry was used instead. Although the values of the scattering cross section and scintillation index will be different in 2 and 3D geometries, the fundamental scattering phenomena, such as Bragg scattering, local tilting, multiple scattering, and shadowing will be present in both 2D and 3D.

Through these numerical experiments, it was found that the broadband scattering strength does not vary as a function of bandwidth for the parameters investigated in this study. The error of this comparison is within Monte-Carlo error. For scintillation index it was found that it becomes greater than one as resolution increases, grazing angle decreases, and spectral strength increases. For larger spectral exponents, the scintillation index is more sensitive to changes in spectral strength and grazing angle.

An overview of the geometry and roughness statistics is presented in Sec.~\ref{sec:geomEnv}. The integral equations and discretization methods are given in Sec.~\ref{sec:integralEquations}, and the incident field in Sec.~\ref{sec:incidentField}. Methods to estimate the broadband scattering cross section and scintillation index are given in Sec.~\ref{sec:estimation}. A discussion on how the parameters of the numerical simulations were selected is given in Sec.~\ref{sec:parameters}. Results are presented in Sec.~\ref{sec:results}, with a discussion and some preliminary hypotheses given in Sec.~\ref{sec:discussion}. Conclusions are given in Sec.~\ref{sec:conclusion}

\section{Geometry and Environment}
\label{sec:geomEnv}
\begin{figure}
  \centering
    \includegraphics[width=3.375in]{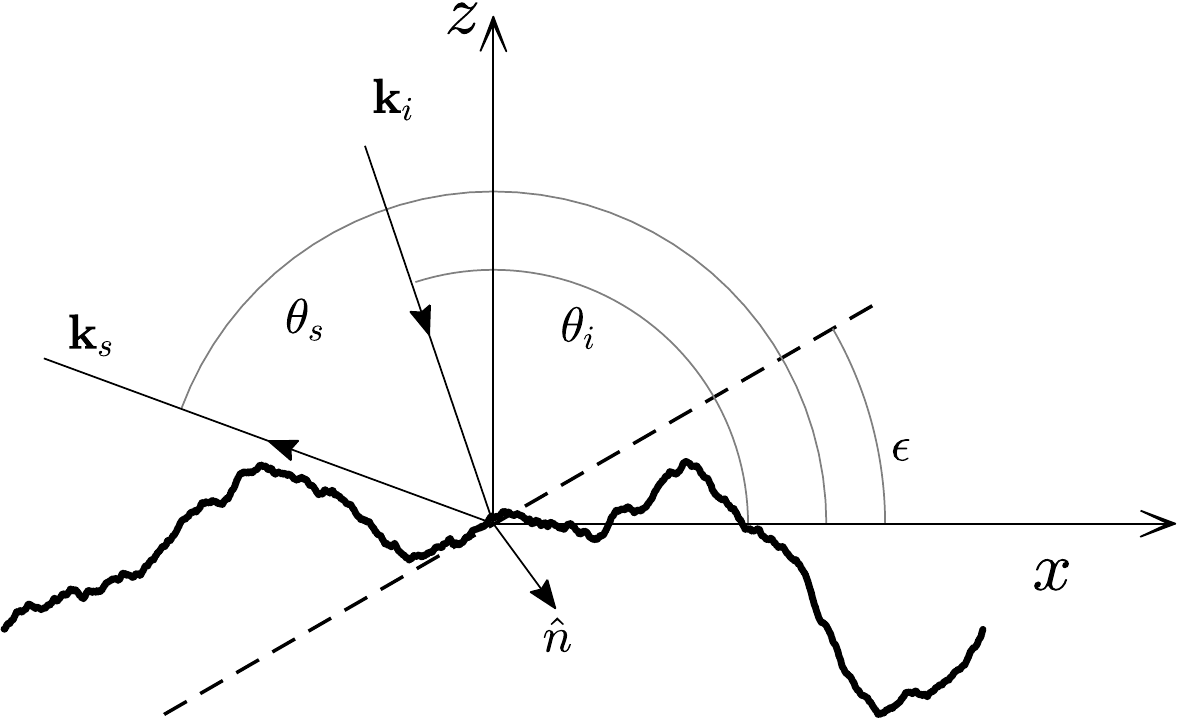}
  \caption{Rough surface scattering geometry. The nominal incident and scattered wave vector are shown, along with the slope angle, $\epsilon$, at the origin, and the unit normal vector at that point, which is defined to point out of the acoustic domain, which is defined as the upper medium.}
  \label{fig:scatteringDiagram}
\end{figure}

The geometry of the scattering problem is presented in Fig.~\ref{fig:scatteringDiagram}. The problem takes place in two dimensions with position vector $\textbf{r}= (x,z)$. The acoustic medium is above the rough surface, defined as $z_s=f(x_s)$ and shown as the thick black line in this figure. The coordinates $(x_s,z_s)$ are points on the rough surface. In this figure, the nominal incident and scattered wave directions are shown with their grazing angles and nominal wave vectors. The slope angle is defined as $\epsilon=\tan^{-1}\left( df(x)/dx \right)$  and is shown in the figure, along with the normal vector pointing out of the acoustic medium (into the lower half-space). The sound speed in the upper medium is $c$, which is taken to be 1500 m/s, but these results can be applied to other sound velocities by performing the appropriate dimensional scaling. The acoustic frequency is $f$, and is related to the wavenumber by $k=2\pi f/c$. Simulations are performed at a center frequency $f_0$, and 3 dB bandwidth $B_{3dB}$. The center wavelength and wavenumber are $\lambda_0$ and $k_0$ respectively

The rough interface is assumed to have wide-sense homogeneity (spatial stationarity) and a Gaussian pdf for both height and slope. By assuming Gaussian statistics, the rough-interface second-order properties can be completely described by its autocovariance function,
\begin{align}
	B(x) = \langle f(y)f(y+x)\rangle
\end{align}
and power density spectrum
\begin{align}
	W(K) = \frac{1}{2\pi}\int B(x) e^{i K x}\, \textrm{d}x.
\end{align}
Several second-order properties of this spectrum are useful for the analysis performed here. In particular, the mean square height, $h^2$ is given by
\begin{align}
	h^2 = \int\limits_{-\infty}^{\infty} W(K)\, \textrm{d} K = B(0).
\end{align}
The mean square slope $s^2$ is
\begin{align}
	s^2 = \int\limits_{-\infty}^{\infty} K^2 W(K)\, \textrm{d} K= -\left. \frac{\partial^2 B(x) }{ \partial x^2 }\right\vert_{x=0}.
\end{align}

The power density spectrum used in this work is the truncated power law,
\begin{align}
	W(K) = \frac{w}{|K|^\gamma}
\end{align}
for $k_{l} \leq |K| \leq k_{u}$, and zero otherwise. The spectral strength is $w$ with units of m$^{3-\gamma}$, and $\gamma$ is the dimensionless spectral exponent. The lower wavenumber cutoff is $k_l=2\pi/L_0$, where $L_0$ is the outer scale. The upper wavenumber cutoff is $k_u=\pi/\ell_0$, where $\ell_0$ is the inner scale. The extra factor of $1/2$ in defining $k_u$ is chosen such that the interval $[-k_u,k_u]$ has a total length of $2\pi/\ell_0$. Random realizations are produced from this power spectrum using the Fourier synthesis technique from \cite{thorsos_1988}, and is given for completeness in Appendix \ref{sec:randomRoughSurfaceGeneration}. The outer scale is specified independently of the surface length, $L$, and is required to satisfy $L_0 < L$. Similarly, the inner scale satisfies $\ell_0 > \delta x $, where $\delta x$ is the sampling interval of the rough interface realization. Although not required in general, the inner scale is smaller than the smallest wavelength with significant energy in these broadband simulations.

For the power-law form used here, the non-dimensional mean square slope and mean square height are
\begin{align}
	s^2 &= \frac{2 k_0^{3-\gamma}w}{3 - \gamma} \left[ \left(\frac{k_u}{k_0}\right)^{3-\gamma} - \left(\frac{k_l}{k_0}\right)^{3-\gamma}\right]\label{eq:rmsSlope} \\
	k_0^2 h^2 &= \frac{2k_0^{3-\gamma} w}{ \gamma-1} \left[ \left( \frac{k_0}{k_l} \right)^{ \gamma-1} - \left( \frac{k_0}{k_u}\right)^{ \gamma-1} \right].\label{eq:rmsHeigh}\, ,
\end{align}
where $k_0$ is the center wavenumber defined in Eq.~(\ref{eq:centerWavenumber}). These parameters have been expressed in a form where the terms outside and inside the brackets parentheses are dimensionless. L'H{\^o}pital's rule can be used to show that the mean square slope is finite for $\gamma=3$, and the mean square height is finite for $\gamma=1$.

The true upper wavenumber is set by the inner scale, $k_u = \pi/\ell_0$. However, the way in which the rough surfaces enter into the acoustical simulations may be subject to an effective upper limit, $k_{u}^\prime=\pi/\ell^\prime$, where $\ell^\prime$ is an effective inner scale. Roughness wavelengths much less than $\lambda_0$ likely have an insignificant effect on the scattered field. Thus, the effective upper wavenumber is likely much less than that defined by the surface sampling. The effective upper limit likely does not affect the scales causing scattering near the Bragg wavelength, but rather sets the upper wavenumber limit for computing the large-scale slope in the slope modulation model examined below. For $\gamma >1$, root-mean-square height is insensitive to the upper cutoff, and more sensitive to the low-wavenumber cutoff. For $\gamma<3$, the root-mean-square slope is sensitive to the upper cutoff, and insensitive to the lower cutoff  (so long as it is sufficiently small). As $k_u$ becomes large, $s$ grows without bound, and an effective upper limit can alleviate this problem. To make the effective upper wavenumber limit explicit, the notation $s_{\ell^\prime}$ is used to denote the rms slope computed using $k_{u}^\prime = \pi/\ell^\prime$.

\section{Integral equations and discretization}
\label{sec:integralEquations}
This study was performed numerically using a discretized form of the 2D Helmholtz-Kirchhoff integral equation for Dirichlet boundary conditions \cite{thorsos_1988}. Although the motivation for this work is seafloor scattering, the assumption of a Dirichlet boundary allows us to focus solely on the role of the rough interface. For a single frequency, this integral equation, defined on the rough interface is \cite{wu_bem},
\begin{align}
	 p_i \left( \textbf{r}_p  \right) = -\int\limits_S \frac{ \partial p \left( \textbf{r}_s \right) }{ \partial n_s} G_k \left(  | \textbf{r}_s - \textbf{r}_p | \right) \textrm{d}S,
	\label{eq:hkie}
\end{align}
where $p_i$ is the incident pressure, $\textbf{r}_p=(x_p,z_p)$ and $\textbf{r}_s=(x_s,z_s)$ are points on the rough surface (with a subscript $s$ denoting the integration variable), $\partial p  / \partial n_s$ is the total pressure normal derivative at $\mathbf{r}_s$, with the normal direction pointing out of the acoustic domain (into the lower space), and
$
G_k(| \textbf{r}_s - \textbf{r}_p |) = (i/4) H_0^{(2)} (k| \textbf{r}_s - \textbf{r}_p |)
$
is the 2D free-space Green function \cite{wu_bem}, where $H_0^{(2)}(z)$ is the zeroth-order Hankel function of the second kind. Note that the Green function used here differs from that used in \cite{thorsos_1988} due to the differing time convention. The normal vector points downward here, as opposed to upward in \cite{thorsos_1988}, although in both cases it points out of the acoustic domain. This integral equation can also describe electromagnetic scattering from 1D corrugated surfaces with perfectly conducting boundary conditions subject to an incident wave with transverse magnetic (also known as p) polarization \cite{toporkov_etal_1998}.

The scattering problem is solved in two steps. First, Eq.~(\ref{eq:hkie}) is numerically solved for $\partial p /\partial n$ on the surface, through discretization of the integral equation using the boundary element method \cite{sauter_schwab,wu_bem}. In particular, piecewise linear basis functions are used to approximate $\partial p/\partial n$, and collocation to compare the true and approximate solution at discrete points. These two steps convert the integral equation into a linear system,
\begin{align}
	V u = b,
\end{align}
where $u$ is the solution vector consisting of the basis function coefficients used in the approximation for $\partial p/\partial n$, and $b =p_i$ evaluated at the discrete collocation points $\textbf{r}_m=(x_m,z_m)$. The matrix $V$ has elements
\begin{align}
	V_{mn} = -\int G_k\left(|\textbf{r}_m -\textbf{r}_s|\right)\phi(\xi_n(\textbf{r}_s))\textrm{d}S.
\end{align}
Here, $\phi(\eta)$ is a linear basis function defined on the interval $\eta \in [-1,\, 1]$. Outside of the interval, $\phi$ is zero. The function $\xi_n$ maps the basis function centered at the $n$-th point from physical space, $\textbf{r}_s$ to the $\eta$ domain. In this case, the basis functions are centered at the same collocation points $\textbf{r}_m$, resulting in a square matrix. Integration is carried out using a 4 point Gauss-Legendre quadrature rule \cite{abramowitz} for nonsingular elements. Due to the weak singularity in the Green function, the diagonal elements of the matrix are computed using a 16 point quadrature rule combined with a variable transformation whose Jacobian exactly cancels the singularity \cite{wu_bem}. LAPACK routines were used to solve the linear system using LU decomposition and back substitution \cite{lapack99}.

Collocation points are defined on the rough surface, $(x_m,z_m)$ with equal spacing, $\delta x$ on the horizontal axis. The method to generate these points is given in Appendix~\ref{sec:randomRoughSurfaceGeneration}. From these points, a cubic approximation is used to construct a continuous and smooth surface. This interpolation process forces the surface normal, and thus $\partial p/\partial n$ to be continuous, which improves the convergence rate of the discretization of the integral operator \cite{atkinson_integral}. The interpolation scheme may extend the region of wavenumber support beyond $\pm\pi/\ell_0$, affecting the estimate of the rms slope and height. However, monotonic piecewise Hermite interpolation \cite{fritsch_carlson_1980} was used, which does not suffer from overshoot and has a negligible effect on the rms slope and height estimates.

Once the surface pressure normal derivative is found, the scattered pressure at a field point in the domain, $p_s(\mathbf{r}_f)$, is found using,
\begin{align}
	p_s(\textbf{r}_f) = \int\limits_S \frac{ \partial p \left( \textbf{r}_s \right) }{ \partial n_s} G_k \left(  | \textbf{r}_s - \textbf{r}_f | \right) \textrm{d}S,
\end{align}
where $\textbf{r}_f = (x_f,z_f)$.
In this work, the field points are equally spaced intervals of one degree at a distance  $R$ from the rough surface, which is approximately 25 times the Rayleigh distance from the surface, $L^2/\lambda_0$, where $L$ is the total surface length and $\lambda_0$ is the center wavelength. This criterion for the far-field is very conservative \cite[Appendix J]{jackson_richardson_2007}, \cite{winebrenner_ishimaru_1986,lysanov_1973}, although it enables the use of asymptotic expansions for the Hankel function. The scattered angle, $\theta_s$ is related to the field point locations using $x_{f} = R \cos\theta_s$, and $z_{f} = R \sin\theta_s$. The variable $p_s(f,\theta_s,\theta_i,R)$ is the complex pressure $p_s$ measured at a location $\mathbf{r}_f$, produced by an incident field with frequency $f$ and nominal incident grazing angle $\theta_i$.

\section{Incident Field}
\label{sec:incidentField}
The incident fields used in this work are broadband pulses whose spatial dependence is an approximation of a plane wave. The nominal directions of the incident and scattered wave vectors are indicated as arrows in Fig.~\ref{fig:scatteringDiagram}. The incident and scattered wave vector lengths vary due to the broadband nature of the field, although the center wave vectors can be defined at the center frequency by the expressions $\textbf{k}_{0i} = (k_{0ix},k_{0iz})$ and $\textbf{k}_{0s}=(k_{0sx},k_{0sz})$. The components are defined in terms of the grazing angles $\theta_i$ and $\theta_s$ (with respect to the horizontal axis) by
\begin{align}
		k_{0ix} &= -k_0 \cos\theta_i &  k_{0sx} &= k_0 \cos\theta_s \\
		k_{0iz} &= -k_0 \sin\theta_i & k_{0sz} &= k_0 \sin\theta_s.
\end{align}
The incident unit wave vector is $\hat{\mathbf{k}}_{0i} = \mathbf{k}_{0i}/k_0$, and the scattered unit vector is $\hat{\mathbf{k}}_{0s} = \mathbf{k}_{0s}/k_0$.

The center wavenumber, $k_0$, is defined by an average of the wavenumber weighted by power spectrum of the transmitted source
\begin{align}
	k_0 =  \frac{\int\limits_{-\infty}^{\infty} \frac{2 \pi f}{c} S^2(f)\, \textrm{d}f}{\int\limits_{-\infty}^{\infty}S^2(f)\, \textrm{d}f},
	\label{eq:centerWavenumber}
\end{align}
where $S(f) = \int s(t)\exp(-i 2 \pi f t) \,\textrm{d}t$ is the linear (amplitude) spectrum of the transmitted pulse, $s(t)$. The transmitted pulse used here was a complex exponential multiplied by a Gaussian envelope, with the time- and frequency-domain forms,
\begin{align}
s(t) =\exp\left(-t^2/\tau^2 + i\omega_0 t\right) \label{eq:gaussianTime}\\
S(f) = \tau \sqrt{\pi} \exp\left(-(f - f_0)^2 \pi^2 \tau^2\right)\, ,
\label{eq:gaussianFreq}
\end{align}
where $\tau$ is a parameter of the pulse length, and $\omega_0=2\pi f_0$ is the center angular frequency. Using the Gaussian form for $S(f)$, the numerator in Eq.~(\ref{eq:centerWavenumber}) is $\pi\sqrt{2\pi}f_0 \tau/c$, and the denominator is $\sqrt{\pi/2}\tau$. The ratio is $2\pi f_0/c = k_0$.

The temporal resolution of the pulse, $\Delta \tau$ is defined by the duration of the pulse envelope between its half power points. For the Gaussian pulse used, this quantity can be obtained by solving the equation $\exp\left(-(\Delta \tau/2)^2/\tau^2\right)= 1/\sqrt{2}$, resulting in $\Delta \tau = \tau\sqrt{2 \ln 2}$. The 3 dB bandwidth, $B_{3dB}$, of the pulse is defined as the full-width, half maximum of $|S(f)|^2$, namely $|S(f_0)|^2/2 = |S(f_0\pm B_{3dB}/2)|^2$. These definitions of the temporal resolution and bandwidth result $B_{3dB} \Delta \tau= 2\ln (2)/\pi \approx 0.44$. For reference, if a rectangular function with full width of $B_{3dB}$ is used for $S(f)$, then $B_{3dB} \Delta \tau\approx 0.88$. The same relationship is obtained if constant envelope pulse of length $\Delta \tau$ is used. Although the rectangular pulse has a larger time-bandwidth product, the Gaussian pulse has no sidelobes in the time-domain, but requires a computational bandwidth much larger than $B_{3dB}$ to approximate a true Gaussian function. The equivalent noise bandwidth (EQNB)\cite{Harris1978} is also needed, which for the Gaussian pulse is $B_{EQNB}=(\tau\sqrt{2\pi})^{-1}$. The constant $A$ is defined in terms of the product
\begin{align}
	\Delta \tau B_{EQNB} = A^{-1}\, ,
	\label{eq:timeBandwidth}
\end{align}
so that $A=\sqrt{\pi/\ln2}$ for the Gaussian pulse. This constant is used later to define an effective inner scale to the rough interface based on its acoustic resolution.

Broadband fields are synthesized from single frequency approximations of a plane wave. This narrowband incident field is the extended Gaussian beam developed by \cite{thorsos_1988} that provides tapering to guard against edge effects entering into the scattering calculation. The form of this field (adapted to this time convention) is given by
\begin{align}
	\begin{split}
	p_i(\textbf{r}_s,f) = p_0 &\exp\left(-i \textbf{k}_i\cdot \textbf{r}_s \left(1 + w_t(\textbf{r}_s)\right)\right) \\
	 \times&\exp\left(-\left(x_s - z_s\cot\theta_i \right)^2/g^2\right),
	\end{split}
	\label{eq:taperedPlaneWave}
\end{align}
where
\begin{align}
	w_t(\textbf{r}_s) = ( k g \sin\theta_i)^{-2} \left[2\left( x_s - z_s \cot\theta_i \right)^2 / g^2 - 1 \right],
\end{align}
$g$ is a width parameter of the incident field, $\mathbf{r}_s$ is a point on the rough surface, and $\theta_i$ is the nominal incident grazing angle. The factor $p_0$ has units of Pa, and is helpful for keeping track of units when estimating the scattering cross section, or broadband cross section. For broadband simulations, Eq.~(\ref{eq:taperedPlaneWave}) is used for each frequency. Since the Gaussian function has an infinite domain of support, it must be truncated to use in numerical simulations.

The function $w_t\left(\textbf{r}\right)$ improves the agreement between the numerical solution of the Helmholtz-Kirchhoff integral equation of the first kind, Eq.~(\ref{eq:hkie}), and the integral equation of the second kind. Discrepancies between these two solutions can result because the incident field satisfies the Helmholtz equation approximately to order $(kg \sin\theta_i)^{-2}$ \cite{thorsos_1988}. Good agreement between the two solutions was observed when $k g\sin\theta_i$ is large. Therefore, to analyze low-grazing angles, (which are important to contemporary synthetic aperture sonar systems, e.g. \cite{fossum_etal_2008,bellettini_pinto_2009,pinto_2011,dillon_2018,sternlicht_etal_2016}), the parameter $g$ must grow as $\theta_i$ approaches zero. This requirement can be thought of as enforcing the constraint that the angular width of the incident beam (full width half max),
\begin{align}
	\Delta \theta =\frac{2\sqrt{2 \ln(2)}}{k g \sin\theta_i},
\end{align}
should be small compared to $\theta_i$. When the relative angular width, defined as $\Delta \theta/\theta_i$ is not small, the direction of the incident field is spread over a large range of angles compared to the incident grazing angle. This situation should be avoided at low grazing angles, since the scattering cross section is commonly rapidly varying in that region.

\section{Estimating time-domain quantities of the scattered field}
\label{sec:estimation}
In this section, expressions are derived for the broadband scattering cross sections and scintillation index of the scattered field due to a broadband incident pulse. The time-domain pressure is computed by
\begin{align}
	p_s(t^\prime,\theta_s,\theta_i,R) = \int\limits_{-\infty}^{\infty} S(f)p_s(f,\theta_s,\theta_i,R)e^{i2\pi f(t^\prime + R/c)}\,\textrm{d}f
	\label{eq:fourierSynthesis}
\end{align}
where $p_s(f,\theta_s,\theta_i,R)$ is the scattered pressure measured at grazing angle $\theta_s$, distance from the origin, $R$, incident field frequency $f$, and nominal incident grazing angle $\theta_i$. The factor $R/c$ in the exponent removes the time delay associated with propagation to the far field when the incident pulse is centered on the origin, so that the scattered pressure time series can be mapped to the mean plane of the rough interface. This delayed time is denoted $t^\prime$, which is used below in the calculation of the effective energy flux. The scattered grazing angle, $\theta_s$ is computed using the location at which the pressure is calculated in the far field, $\theta_s = \tan^{-1}(z_f/x_f)$. In practice the integral is computed using the fast Fourier transform.

The broadband scattering cross section, $\sigma_{bb}$, is computed directly from the scattered pressure in the time domain. Although there is no general framework for this kind of quantity, motivation is provided here by comparison to the definition of the scattering cross section in the frequency domain. In practice, finite resolution in scattering measurements is sometimes obtained using short pulses \cite{urick_1954,urick_underwater_sound}, and this definition accords with the sonar equation used in those situations.

The frequency domain version of the cross section, $\sigma_f$ for a monochromatic plane wave from a rough interface of length $L$, was given in Sec.~\ref{sec:introduction} as Eq.~(\ref{eq:planeWaveSigmaDefinition}), due to \cite{thorsos_1988}. This expression can be cast in terms of the incident energy-flux \cite[p. 255]{landau_fluids},\cite{thorsos_1988} passing through the plane $z=0$
\begin{align}
	\sigma_f = \frac{\langle I_s \rangle R \sin\theta_i}{E_f},
	\label{eq:sigmaEnergyFlux}
\end{align}
where
\begin{align}
	E_f = \int\limits_{-\infty}^{\infty} \left[ \left. \frac{1}{2} \mathrm{Re}\lbrace p_i \mathbf{v}_{i}^\ast \rbrace \cdot \hat{n}\right. \right|_{z=0}\, \mathrm{d}x ,
\end{align}
is the total energy flux passing through the $z=0$ plane, $p_i$ is the incident complex pressure, $\mathbf{v}_{i}$ is the incident acoustic particle velocity, and $\hat{n}$ is the unit normal vector of the line $z=0$. Since the energy is directed downards towards the rough interface, $\hat{n}$ was chosen to be $-\hat{z}$. In \cite{thorsos_1988}, the incident acoustic energy was directed upwards towards the rough interface, and correspondingly, their normal vector pointed up. Making this substitution, the energy flux is
\begin{align}
	E_f = -\int\limits_{-\infty}^{\infty} \left[ \left.\frac{1}{2} \mathrm{Re}\lbrace p_i {v}_{iz}^\ast \rbrace\right. \right|_{z=0}\, \mathrm{d}x ,
	\label{eq:energyFluxIntegral}
\end{align}
where $v_{iz}$ is the vertical component of the incident acoustic particle velocity. For an untapered plane wave incident on a rough surface of length $L$, $E_f=I_i L \sin\theta_i$, and the definition in Eq.~(\ref{eq:planeWaveSigmaDefinition}) is recovered.

This definition can be extended to an incident plane wave with a time dependence set by $e^{i\omega_0 t}$ times a rectangular pulse of length $\tau$. Here, a new quantity is defined called the broadband scattering cross section, $\sigma_{bb}$. It might seem reasonable to start with Eq.~(\ref{eq:sigmaEnergyFlux}) with $E_f$ evaluated in the time-domain. In this case, $E_f=I_i c \tau\sin  \theta_i/\cos\theta_i$, and thus $\sigma_{bb} = (\langle I_s \rangle R)/(I_i c\tau/\cos\theta_i)$. However, this definition is insufficient because in the case of an incident pulse, the (effective) ensonified length that is relevant for the scattering cross section is set by both the incoming and outgoing angles. Physically, the dependence on the scattered field point (or angle) is due to the fact that at a given instant in time, only part of the ensonified surface contributes to the total field measured at a point in the far field.

To account for this effect, the vertical component of the energy flux density can be written as a function of space and time
\begin{align}
	e_f(t,\textbf{r}) = \frac{1}{2}Re\lbrace p_i v_{iz}^\ast \rbrace\, ,
	\label{eq:effectiveEnergyFluxDensity}
\end{align}
which is the integrand of (\ref{eq:energyFluxIntegral}). Instead of integrating this quantity over $x$ to obtain the energy flux, a delayed version of $e_f$ is integrated over $x$ with $z=0$ to obtain the effective energy flux. This quantity is defined as
\begin{align}
	E_f^{\prime} = -\int\limits_{-\infty}^{\infty} e_f(t - t_s,\mathbf{r})\vert_{z=0} \,\mathrm{d}x\, ,
	\label{eq:modifiedEnergyFlux}
\end{align}
where
\begin{align}
	t_s = |\mathbf{r}_f - \mathbf{r}|/c \approx (R - x \cos\theta_s - z\sin\theta_s )/c\, ,
\end{align}
and $R = |\mathbf{r}_f|$. This time-delay takes into account the alteration in the region that contributes to the scattered field at a given instant in time due to the position of the field point. For a line source, or other compact source configuration, a similar analysis can be performed by taking into account the constant-time (isochronous) ellipse for the transmitted pulse and geometry.

The broadband scattering cross section is defined in terms of the effective incident energy-flux, analogous to the frequency-domain version
\begin{align}
	\sigma_{bb} = \frac{\langle I_s \rangle R \sin\theta_i}{E_f^\prime}\, .
	\label{eq:sigmaBBmod}
\end{align}
Note that $E_f^\prime$ may in general be a function of time, since the pulse may be attenuated by a spatial taper (as it is here), or by the beam pattern of a transducer in experiments. To use the entire time series scattered by the rough interface due to the broadband pulse, the energy flux must be calculated as a function of time. For a plane-wave (untapered) pulse the energy flux is independent of time, for all pulse shapes.

First, this defintion will be demonstrated for a plane wave with a rectangular pulse shape, to build intuition. The incident pressure in this case is given by
\begin{align}
	p_i(\mathbf{r},t) = p_0 e^{i\omega_0 (t - \hat{\mathbf{k}}_i \cdot \mathbf{r}/c)}  \Pi\left(\frac{t - \hat{\mathbf{k}}_i\cdot \mathbf{r}/c}{\tau/2} \right)
\end{align}
where $\mathbf{r}$ is a general point in space, $\hat{\mathbf{k}}_i$ is the incident unit wave vector, $\Pi(x)=1$ if $|x| \leq 1$ and zero otherwise, and sets the incident pulse duration. The effective energy flux is then $E_f^\prime =I_i c \tau \sin\theta_i /\nu_x$, where $\nu_x = \cos\theta_i + \cos\theta_s$. For a broadband rectangular pulse, the sine factors cancel, and
\begin{align}
	\sigma_{bb} = \frac{\langle I_s \rangle R}{I_i c\tau/\nu_x}.
\end{align}
The denominator now includes dependence on the scattered direction, and contains the effective ensonified length, $L_{eff}$. The form $L_{eff} = c\tau/\nu_x$ is consistent with the down-range resolution of bistatic synthetic aperture radar, given in \cite{moccia_renga_2011} after converting between different angle conventions.

The effective energy flux for a broadband Gaussian pulse and incident Gaussian beam is derived in Appendix \ref{sec:energyFluxAppendix}. The result is
\begin{align}
	E_f^\prime &= \frac{|p_0|^2}{2\rho_0 c} \sin\theta_i L_{eff} D(t^\prime) \label{eq:ef_simplified}\\
		L_{eff} &= \frac{\sqrt{ \pi / 2 }} { \sqrt{ g^{-2} + (c\tau/\nu_x)^{-2}}}\label{eq:Leff} \\
		D(t^\prime) &= e^{-\frac{2 (ct^\prime/\nu_x)^2}{g^2 + (c\tau/\nu_x)^2}} (1 + J_1(1 + J_2 J_3))\, . \label{eq:directivity}
\end{align}
where
\begin{align}
		J_1 &= \frac{\cot\theta_i\nu_x}{ \sin\theta_i g^2(2g^{-2} + (c\tau/\nu_x)^{-2})} \label{eq:J1}\\
		J_2 &= \frac{2\left(g^{-2} +(c\tau/\nu_x)^{-2} \right)}{\left(2g^{-2} +(c\tau/\nu_x)^{-2}\right)} \label{eq:J2} \\
		J_3 &= \frac{\left(\frac{t^\prime}{\tau}\right)^2\left(1 - \chi^2\right) - \left(\frac{\omega_0 \tau}{2} \right)^2 \left(1 + \chi^2\right)}{\left(\frac{t^\prime}{\tau}\right)^2\left(1 - \chi^2\right)^2 - \left(\frac{\omega_0 \tau}{2} \right)^2 \left(1 + \chi^2\right)^2} \label{eq:J3} \\
		\chi &= \frac{\nu_x/(c\tau)}{\sqrt{2g^{-2} + (c\tau/\nu_x)^{-2}}} \label{eq:chi}\, .
\end{align}
$L_{eff}$ is the effective ensonified length, and $D(t^\prime)$ can be thought of as the directivity function that captures the effect of the incident pulse traveling throughout space and changing amplitude due to the incident beam. Away from specular, if $g \gg c\tau/2$, then the $L_{eff} \approx (\pi/2)^{1/2}c\tau/\nu_x$. Note that when $\sin\theta_i g \nu_x / (c\tau) \gg 1$, $J_1$ is small, $J_2$ is of order unity, and $J_3$ tends to 1/2.

The broadband scattering cross section is defined here in terms of an intermediate variable, the unaveraged broadband scattering cross section as a function of time, incident angle and scattered angle, $q(t^\prime,\theta_s,\theta_i)$,
\begin{align}
	q(t^\prime,\theta_s,\theta_i) = \frac{|p_s(t^\prime,\theta_s,\theta_i,R)|^2 R}{|p_0|^2 L_{eff} D(t^\prime)}.
	\label{eq:timeDomainScsatteringCrossSection}
\end{align}
To reduce uncertainty, $q(t^\prime ,\theta_s,\theta_i)$ is computed for different roughness realizations, and results concatenated. The broadband scattering cross section can be computed by averaging $q$ over time, $t^\prime$ and random realizations, $N_E$.
\begin{align}
	\sigma_{bb}(\theta_s,\theta_i) = \langle q(t^\prime,\theta_s,\theta_i) \rangle_{t^\prime,N_E}.
	\label{eq:timeDomainSigmaDefinition}
\end{align}
In Sec.~\ref{sec:results}, the broadband cross section is compared to the frequency domain scattering cross section, $\sigma_f$, calculated using Eq.~(14) from \cite{thorsos_1988}. Although the scattering cross section is exclusively a frequency-domain quantity, the subscript $f$ is included for clarity.

\begin{figure}
	\centering
	\includegraphics[width=3.375in]{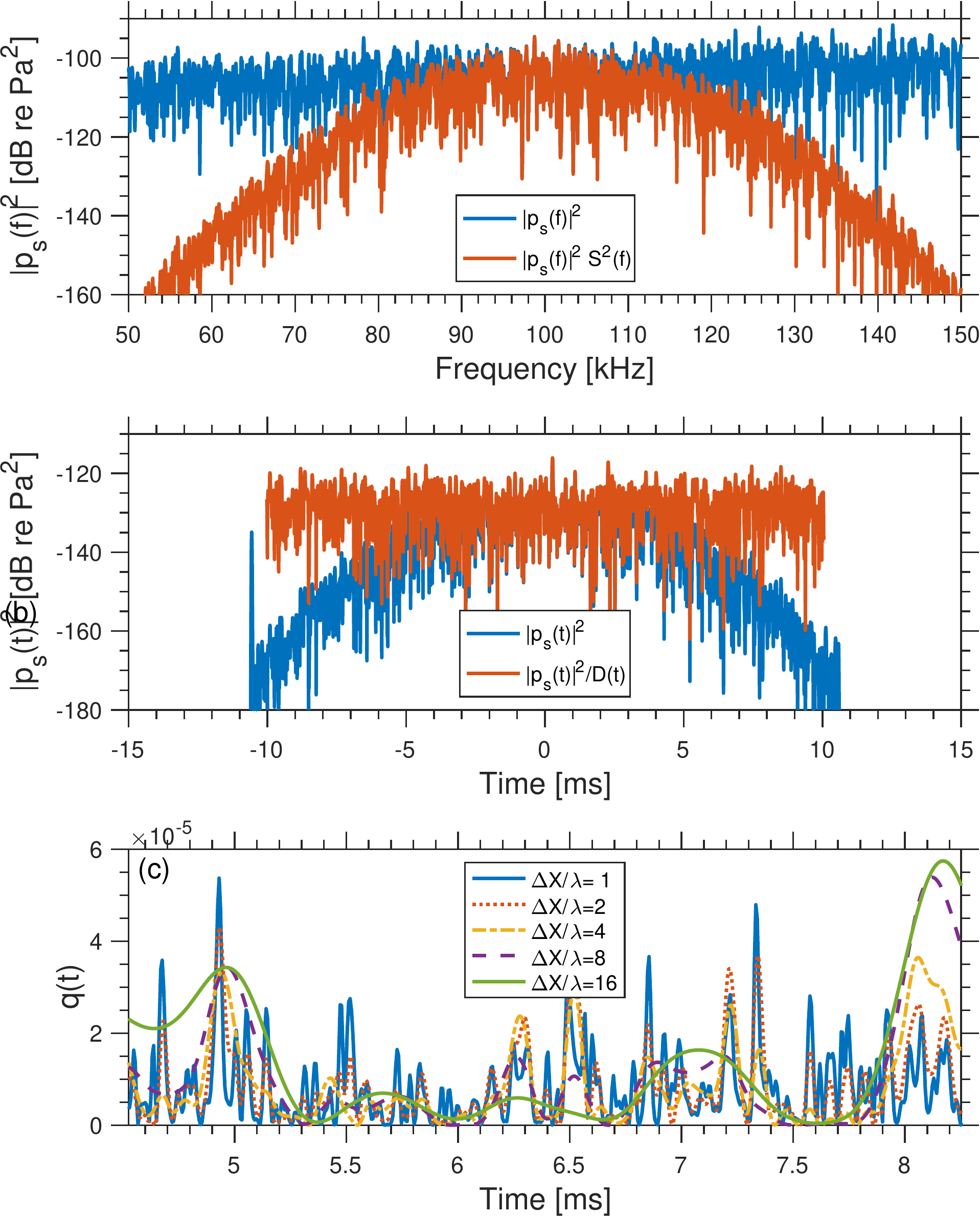}
	\caption{(color online) Steps to estimate broadband scattering  cross section. The frequency domain scattered pressure, and the pressure weighted by the source spectrum are given in (a) (angle arguments to $p_s$ have been omitted). In $(b)$ the raw time-domain scattered-pressure magnitude-squared is plotted, along with a version divided by $D(t^\prime)$ (defined in Eq.~(\ref{eq:directivity})). In (c), the time-domain unaveraged broadband scattering cross section (defined in Eq.~(\ref{eq:timeDomainScsatteringCrossSection})) is plotted for five different pulse lengths. The incident and scattered grazing angles were 20$^\circ$, $g$ was 3.75 m, and $R$ was placed conservatively in the far-field of the rough interface.}
	\label{fig:dataProcessingSteps}
\end{figure}

An example realization of the scattered pressure in the frequency domain is plotted in Fig.~\ref{fig:dataProcessingSteps}(a). The frequency domain pressure is plotted as the raw scattered pressure, and also weighted by the amplitude spectrum, $S(f)$. The time-domain pressure-squared after weighting by $S(f)$ and an inverse Fourier transform, is plotted in Fig.~\ref{fig:dataProcessingSteps}(b). The time-domain squared pressure magnitude contains fluctuations with a characteristic time-scale close to the pulse length, as well as deterministic changes due to the incident beam used in the Helmholtz integral calculations. One example of $q(t^\prime,20^\circ,20^\circ)$ is plotted in Fig.~\ref{fig:dataProcessingSteps}(c) for five different pulse resolutions.

The mapping $x = -ct^\prime/\nu_x$, can be used to convert the time series of the scattered pressure to the mean plane. This mapping gives the location of the center of the incident pulse on the mean plane, as is sometimes performed for imaging sonars. If single-scattering is assumed, then the time-domain scattered field can be mapped to the horizontal position of the rough interface. If the surfaces are very rough and multiple scattering is present, then the scattering will occur from locations other than $x=-ct^\prime/\nu_x$.

The scintillation index, $SI$ is also examined. It is the variance of the scattered intensity divided by the square of the mean scattered intensity \cite[p. 437]{ishimaru_volII}. Since $SI$ is invariant under a multiplication of the intensity by a constant, the unaveraged broadband cross section, $q$, may be used instead of the scattered intensity, so that
 \begin{align}
	SI = \frac{\langle q^2\rangle - \langle q\rangle^2}{\langle q\rangle^2}\,.
\end{align}
The scintillation index characterizes the fluctuations in the scattered field. If $SI=1$, then the magnitude of the complex pressure (known as the envelope) has a Rayleigh distribution and its real and imaginary components are Gaussian. If $SI>1$, then the pdf of the scattered field is heavy-tailed, which means that there is a higher probability of occurrence of high-amplitude events compared to the Rayleigh distribution.
\section{Parameters for Numerical Experiments}
\label{sec:parameters}
\subsection{Signal Parameters}
The objective of this work is to study the resolution (or bandwidth) dependence of the scattered field. These experiments covered the resolutions typically used in narrowband scattering experiments \cite{jackson_etal_1986,williams_etal_2002}, with the resolution cell on the order of 10 or more wavelengths, down to a value of one wavelength, which is on the order of what is achievable by modern SAS systems. Specific values of $\Delta X/\lambda_0 = (1,2,4,8,16)$ were used. The proportional spatial resolutions correspond to temporal resolutions $\Delta \tau f_0 = (2,4,8,16,32)$ at small grazing angles, since $\Delta X = c \Delta \tau/(2 \cos \theta_i)$ for backscattering. In all cases, the resolution is defined for $\theta_i = \theta_s = 0$. At larger incident and scattered grazing angles, the resolution will be somewhat larger than the values listed in the figures and tables presented below.

High-frequency acoustic imaging systems provided the motivation for this work, and thus the simulations used a center frequency of 100 kHz. However, the parameters of the simulation were specified in a non-dimensional fashion. As long as every dimensional quantity is scaled properly, results of these simulations should be valid for lower frequencies with larger roughness parameters, and longer surface lengths. To check whether the non-dimensional scaling was valid, one of the simulations was performed at 10 kHz as well, and the roughness parameters were scaled accordingly. A sound speed of 1500 m/s was used for all simulations.

\subsection{Roughness parameters}
Roughness parameters were specified using two dimensional constants, the center wavenumber, $k_0$ (through a combination of $f_0$ and $c$) and spectral strength, $w$. Three spectral exponents were used, $\gamma=(1.5,2,2.5)$, since this parameter has been observed to vary for measured seafloor roughness \cite[Ch 6]{jackson_richardson_2007}. A spectral exponent of 2 was used for both 100 kHz and 10 kHz. The $\gamma=1.5$ and $\gamma=2.5$ simulations used a center frequency of 100 kHz only.

It now remains to specify the spectral strength. For $\gamma=2$, spectral strengths of $w_{2} = (1\times10^{-6}, 1\times10^{-5}, 2\times10^{-5}, 3\times10^{-5}, 4\times10^{-5})$ m were used, where a subscript on the spectral strength denotes that it is used for a specific value of the spectral exponent. These values resulted in  $SI\approx1$ for the smallest $w_2$, and $SI>1$ for larger values. These values are much smaller than the measurements with exponent close to 2 summarized in Table 6.1 of \cite{jackson_richardson_2007}. The same is true for spectral strength for other $\gamma$ in Table~\ref{tab:roughnessParameters}.

Given our interest in investigating the local tilting hypothesis, the spectral strength for other values of $\gamma$ were specified such that they resulted in equal rms slope when the effective inner scale was held constant at $\ell^\prime=A\lambda_0 \approx 2\lambda_0$, where $A$ is defined in Eq.~(\ref{eq:timeBandwidth}). The factor of $A$ is included since $(A\Delta X)$ is the length scale relevant for computing rms slope and height with an effective upper wavenumber limit. With this requirement, values of $w$ for other $\gamma$ were computed using
\begin{align}
	w_{\gamma}= w_{2} (3 - \gamma) \frac{k_{u}^\prime- k_l^\prime}{k_{u}^{\prime 3 - \gamma} - k_l^{\prime 3 - \gamma}}.
	\label{eq:consistentW}
\end{align}
The effective upper cutoff was set to $k_{u}^\prime=k_0/(2A)$, since $\ell^\prime = A\lambda_0$.

The true inner scale was $\ell_0=\lambda_0/6$ to ensure that the Bragg condition was satisfied for all wavelengths in the broadband simulation with significant support. This criterion ensured that $1.5 f_0$ was the largest frequency that satisfied the Bragg condition. The pulse with the largest bandwidth has a spectrum that is 67 dB down at $1.5f_0$ compared to the the peak value at $f_0$. The outer scale was $L_0=112.5\lambda_0 = L/10$ to ensure that a broad range of scales were included in the power law spectrum, but the surface length was sufficiently larger than the outer scale.

Roughness parameters for the 100 kHz simulations are summarized in Table \ref{tab:roughnessParameters}. Root-mean-square slope is given in degrees for different values of $\ell^\prime$, the effective inner scale. With $\ell^\prime=\ell_0=\lambda_0/6$, the true rms slope of the rough interface is given. Values of $s_{\ell^\prime}$ for $\ell^\prime$ between $A\lambda_0$ and $16A\lambda_0$ are also given, each of which corresponds to the acoustic resolutions used in the numerical simulations. For the $\gamma=2$ case, the true rms slope varied between $2.9^\circ$ and $18^\circ$, depending on the spectral strength. For the $\gamma=1.5$ case the true rms slope varied between $5.3^\circ$ and $31^\circ$, and for the $\gamma=2.5$ case it varied between $1.6^\circ$ and $10^\circ$. The effective rms slopes were much smaller than the true values. The large values of true rms slope result because this quantity increases without bound as the inner scale tends to zero. For the 10 kHz simulations, the same $s_{\delta x}$ and $kh$ were chosen. This condition can be satisfied if the spectral strength, surface length, inner scale, and outer scale, for $100$ kHz and $\gamma=2$ are all multiplied by 10. For other values of $\gamma$, a different scaling must be used.

\begin{table}
\singlespacing
\centering
\begin{adjustbox}{width=3.375in}
\begin{tabular}{l| l l l l l l l l}
\hline \hline
$\gamma$ & $w$ & $s_{\lambda_0/6}$ &$s_{A\lambda_0}$ & $s_{2A\lambda_0}$& $s_{4A\lambda_0}$ & $s_{8A\lambda_0}$& $s_{16A\lambda_0}$& $k_0h$ \\
- & m$^{3-\gamma}$ & $^\circ$ & $^\circ$ & $^\circ$ & $^\circ$ & $^\circ$ & $^\circ$ & - \\
\hline
\multirow{5}{*}{2} %
& 1.00$\times10^{-6}$ & 2.87 & 0.79 & 0.55 & 0.37 & 0.24 & 0.13 & 0.31\\
& 1.00$\times10^{-5}$ & 9.00 & 2.49 & 1.73 & 1.17 & 0.75 & 0.40 & 0.97\\
& 2.00$\times10^{-5}$ & 12.62 & 3.52 & 2.44 & 1.66 & 1.06 & 0.56 & 1.37\\
& 3.00$\times10^{-5}$ & 15.33 & 4.31 & 2.99 & 2.03 & 1.30 & 0.69 & 1.68\\
& 4.00$\times10^{-5}$ & 17.57 & 4.97 & 3.45 & 2.34 & 1.50 & 0.80 & 1.94\\
 \hline
\multirow{5}{*}{1.5} %
& 1.47$\times10^{-7}$ & 5.34 & 0.79 & 0.47 & 0.27 & 0.15 & 0.07 & 0.22\\
& 1.47$\times10^{-6}$ & 16.44 & 2.49 & 1.47 & 0.86 & 0.48 & 0.23 & 0.71\\
& 2.93$\times10^{-6}$ & 22.65 & 3.52 & 2.08 & 1.21 & 0.68 & 0.32 & 1.00\\
& 4.40$\times10^{-6}$ & 27.07 & 4.31 & 2.55 & 1.49 & 0.83 & 0.39 & 1.23\\
& 5.86$\times10^{-6}$ & 30.54 & 4.97 & 2.94 & 1.72 & 0.96 & 0.45 & 1.42\\
\hline
\multirow{5}{*}{2.5} %
& 5.90$\times10^{-6}$ & 1.61 & 0.79 & 0.63 & 0.48 & 0.35 & 0.21 & 0.44\\
& 5.92$\times10^{-5}$ & 5.09 & 2.49 & 1.99 & 1.53 & 1.11 & 0.65 & 1.39\\
& 1.18$\times10^{-4}$ & 7.18 & 3.52 & 2.81 & 2.17 & 1.57 & 0.93 & 1.96\\
& 1.78$\times10^{-4}$ & 8.77 & 4.31 & 3.44 & 2.66 & 1.92 & 1.13 & 2.41\\
& 2.37$\times10^{-4}$ & 10.11 & 4.97 & 3.97 & 3.07 & 2.21 & 1.31 & 2.78\\
\hline \hline
\end{tabular}
\end{adjustbox}
\caption{Roughness parameters used in the simulations. All parameters are listed for $f_0=100$kHz and $c=1500$ m/s. Simulations at 10 kHz use the same dimensionless mean square parameters. Units for each parameter are given in the second row. The rms slope is reported using the upper limit computed with the sampling interval for the rough surface, as well as using the acoustic resolution. The parameter $A\approx2.12$.}
\label{tab:roughnessParameters}
\end{table}

\begin{table}
	\centering
	\begin{tabular}{l | r r r}
	\hline \hline
	$\Delta X/\lambda_0$ & $N_p$ & Err (\%) & Err (dB) \\
	1 & 4275 & 1.53 & 0.066 \\
	2 & 2137 & 2.16 & 0.093 \\
	4 & 1069 & 3.06 & 0.13 \\
	8 & 534 & 4.33 & 0.18\\
	16 & 267 & 6.12 & 0.26 \\
	\hline \hline
	\end{tabular}
\caption{Number of independent samples, $N_p$ for the different resolutions at small grazing angles, and relative uncertainty associated with the finite ensemble assuming intensity is exponentially distributed. Uncertainty is reported in terms of percent and decibels. The frequency-domain simulations used 5000 surface realizations, with an uncertainty of 1.14\%, or 0.061 dB.}
\label{tab:numberOfIndependentSamples}
\end{table}
\subsection{Sampling parameters}
\label{sec:samplingParameters}
The sampling interval, $\delta x$ was specified so that the errors caused by discretization of the integral equation were not observable to within Monte-Carlo error. Since very large numbers of independent samples were used, these estimates of the broadband scattering cross section had uncertainty of about 0.07-0.3 dB. With this small uncertainty, a noticeable bias in the scattering strength occurred when $\delta x/\lambda_0>1/12$, due to discretization error. A conservative value of $\delta x/\lambda_0=1/15$ was therefore used to minimize this bias.

The focus in this work is moderate to low grazing angles. The lower limit to the grazing angles that can be reliably estimated in these numerical simulations is set by the surface length, which in turn is set by the memory limitations and acceptable number of CPU hours used. These latter constraints limited us to $g=250\lambda_0$. At the center frequency, the relative angular width for that value of $g$ was about 3.5\% at 10$^\circ$ grazing angle. Lower frequencies are more problematic, since, for a constant value of $g$, decreasing the frequency will increase the angular width of the field. At the lower 6 dB down point of the largest bandwidth signal used (equivalent to $0.844 f_0$), the angular width for $g=250\lambda_0$ was approximately 4.2$^\circ$ at a nominal angle of $10^\circ$ grazing angle. With these angular widths, $10^\circ$ was taken to be an acceptable, if conservative, lower limit to the grazing angles that can be reliably estimated in this work. In choosing the surface length, $g$ was increased to $400\lambda_0$ without any change in the behavior of $\sigma_{bb}$ or $SI$ above 10$^\circ$. Grazing angles less than $10^\circ$ likely require fast approximate methods to solve the integral equation, such as the fast multipole method \cite{liu_bem_fmm}, or hierarchical matrices \cite{hackbusch_2015}.

The rough surface length, $L$ was set to $4.5g$ to allow the incident field taper to decay sufficiently at the edges of the computational domain. Near the edges, 2.5\% of the time-domain samples were discarded on each side to remove edge effects (e.g. the large peak at -11 ms in the blue curve in Fig.~\ref{fig:dataProcessingSteps}(b)), so that only 95\% of the full time series available was used.

In the broadband simulations, four roughness realizations were used, and 5000 realizations used in the frequency domain. For the broadband case, the time-domain response for each angle contains many independent samples, $N_p$ (including multiple realizations). At small grazing angles for all ensembles, $N_p$ is $4\times0.95 L/(\Delta X)$, where the factor of $0.95$ results from discarding samples near the edges of the surface. Assuming Gaussian pressure statistics (or exponentially distributed intensity statistics), one can define a standard error $Err_{\mathrm{std}}= 1/\sqrt{N_p}$ which characterizes the relative uncertainty of the broadband scattering cross section estimate. The decibel representation is $Err_{\mathrm{dB}}=10 \log_{10}(1 + 1/\sqrt{N_p})$. However, the scattered intensity is manifestly not exponentially distributed for large values of spectral strength, small pulse length, and small grazing angles. A more realistic characterization of the uncertainty is to divide the standard deviation of the intensity by the mean intensity and $\sqrt{N_p}$, amounting to $\sqrt{SI/N_p}$. Since the scintillation index is shown in the next section to vary with roughness parameters and grazing angle, it is difficult to give an overall characterization of the uncertainty. Therefore, uncertainty calculated using Gaussian statistics are given in Table ~\ref{tab:numberOfIndependentSamples}, with the smallest error of 1.5\% or 0.066 dB, and largest uncertainty of 6.1\% or 0.26 dB. If the uncertainty for the non-Gaussian cases are desired, then the scintillation index plots presented in Figs.~(\ref{fig:SI_100kHz_gamma2}-\ref{fig:SI_100kHz_gamma2.5}) can be used along with $N_p$ to provide this quantity.

The frequencies required for the largest bandwidth simulation $\Delta \tau f_0=2$, spanned approximately $0.2f_0$ to $1.8f_0$. This computational bandwidth was about seven times the largest 3 dB bandwidth used for the Gaussian spectrum. Using a frequency spacing of $\delta f =c/(3L)$, the number of frequencies per simulation was approximately 6000. For the proportional bandwidths studied in this work, these surface parameters resulted in surfaces with $N=16,875$ points. The matrix $V$ resulting from this discretization required 4.3 GB of memory storage for double precision complex numbers. Simulations were performed on the Hamming high-performance computing cluster at the Naval Postgraduate School.

\section{Results}
\label{sec:results}
\begin{figure}
	  \centering
	    \includegraphics[width=3.375in]{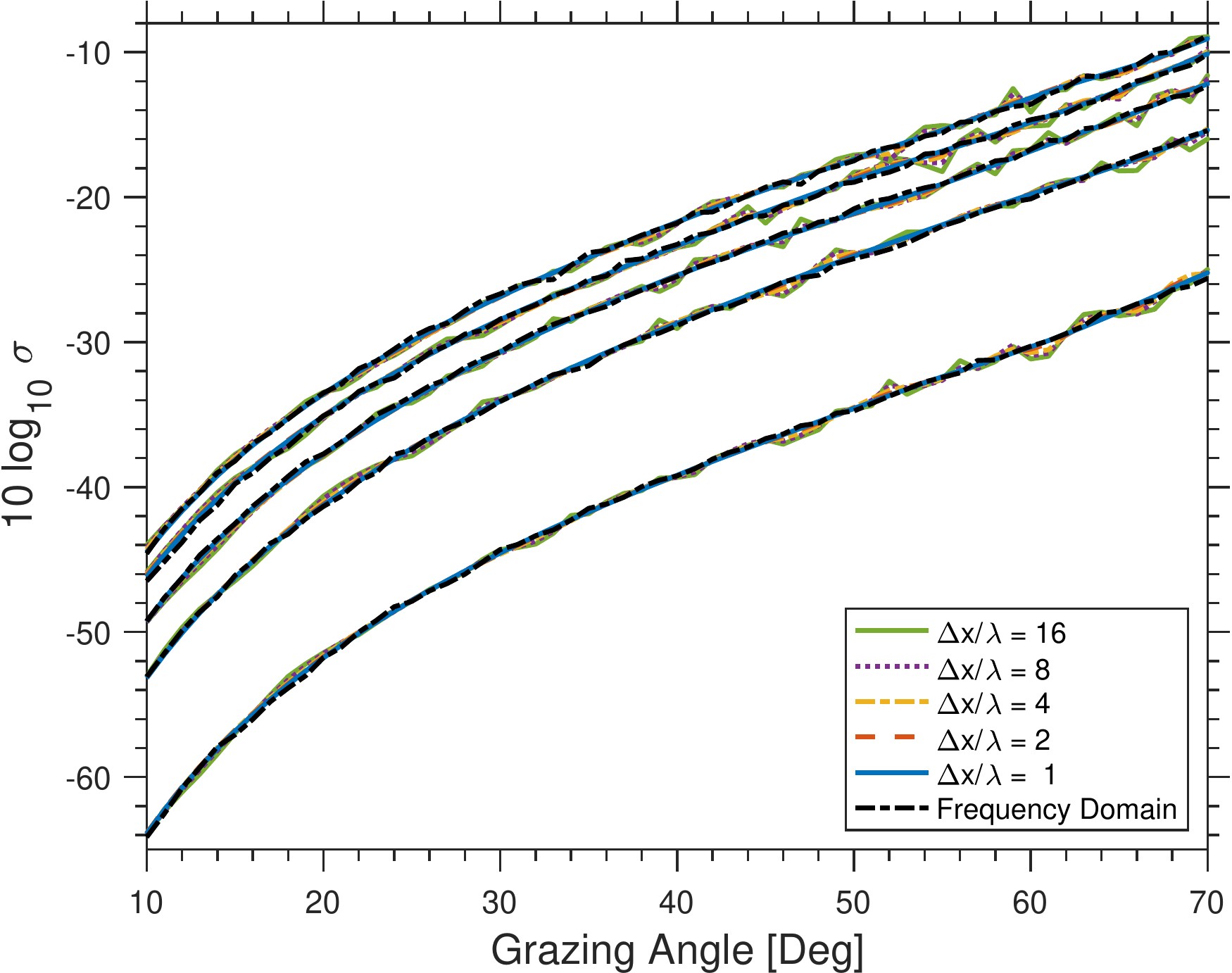}
	  \caption{(color online) Scattering Strength comparison for $f_0$= 100 kHz and $\gamma=2$. Each color represents scattering strength estimated using different bandwidths, as indicated by the caption. The dot-dashed line indicates the frequency-domain estimate. Five different values of spectral strength are plotted on separate lines, with scattering strength monotonically increasing with spectral strength (i.e. the smallest spectral strength has the lowest scattering strength). Values for the spectral strength are given in Table~\ref{tab:roughnessParameters}. Note that the random realizations are generated only with inner cutoff scale of $\ell_0 = \lambda_0/6$. The time-domain results are difficult to distinguish, and the ratio for the time-domain and frequency results are shown in Fig.~\ref{fig:SSError_100kHz_gamma2}.}
	  \label{fig:SS100kHz_gamma2}
\end{figure}

\begin{figure}
  \centering
    \includegraphics[width=3.375in]{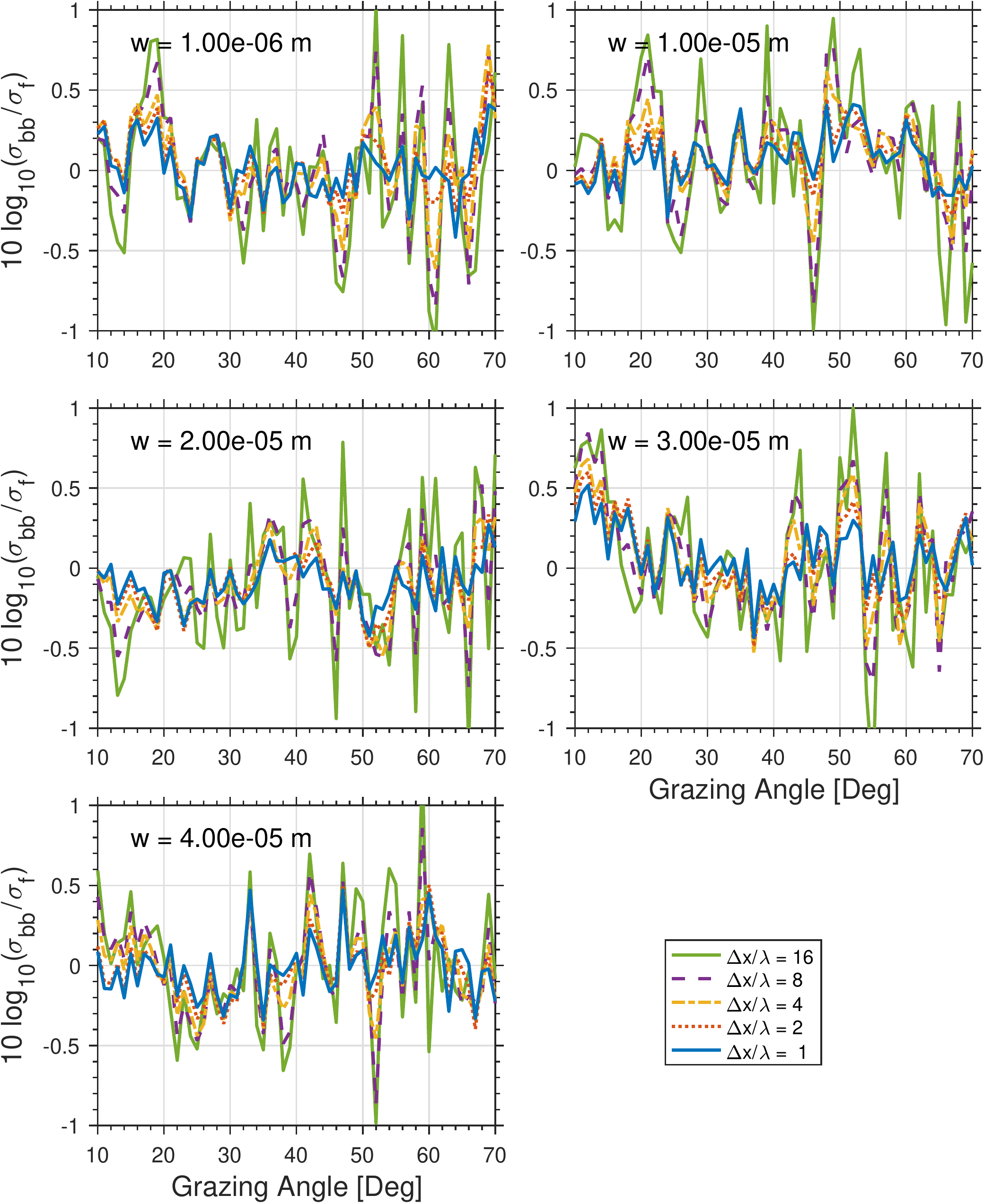}
  \caption{(color online) Scattering Strength ratio for $f_0$=100 kHz and $\gamma=2$}
  \label{fig:SSError_100kHz_gamma2}
\end{figure}

\begin{figure}
  \centering
    \includegraphics[width=3.375in]{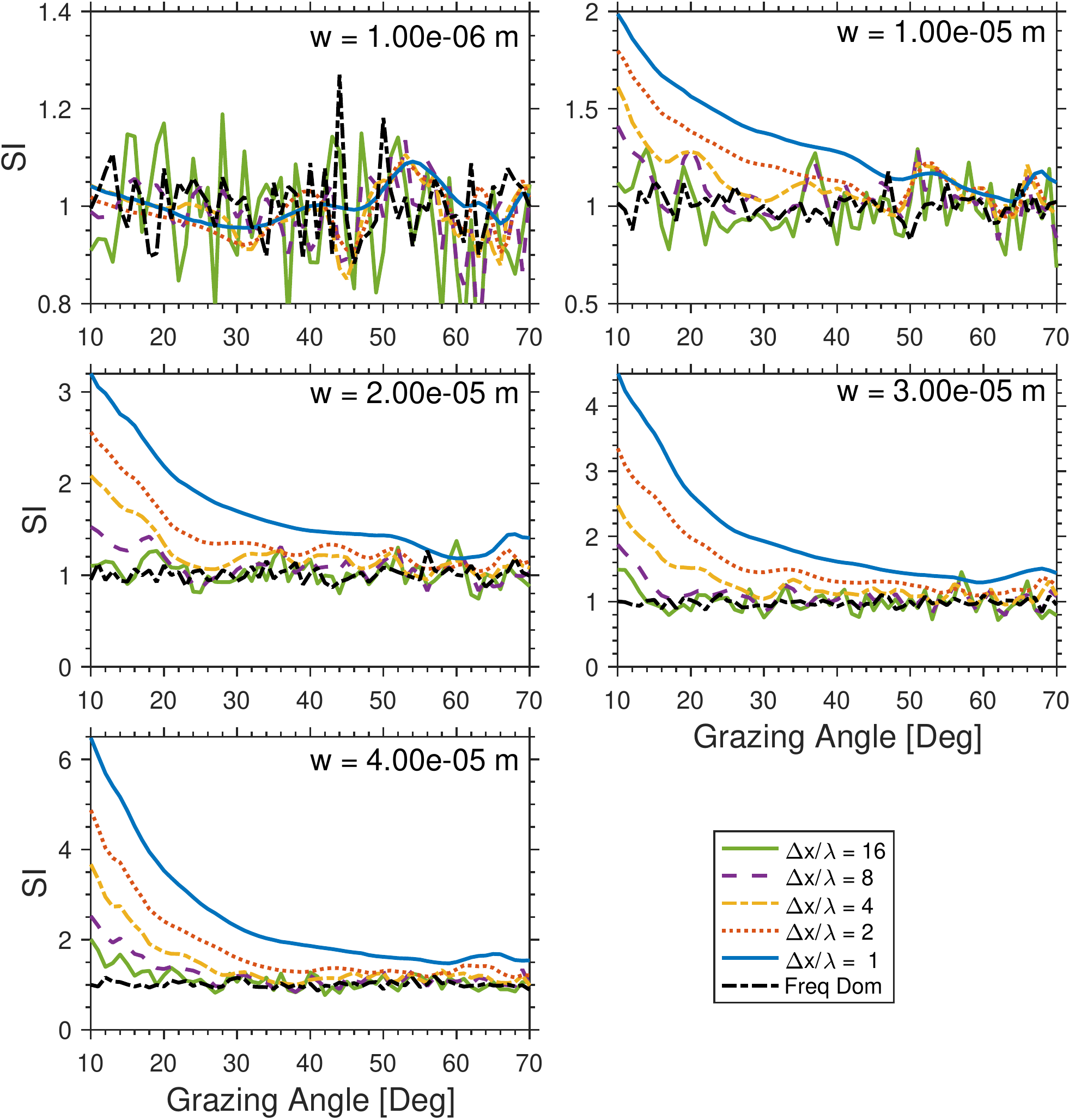}
  \caption{(color online) Scintillation Index for $f_0$=100 kHz and $\gamma=2$. Note that the vertical axis is different for each subfigure.}
  \label{fig:SI_100kHz_gamma2}
\end{figure}

\begin{figure}
  \centering
    \includegraphics[width=3.375in]{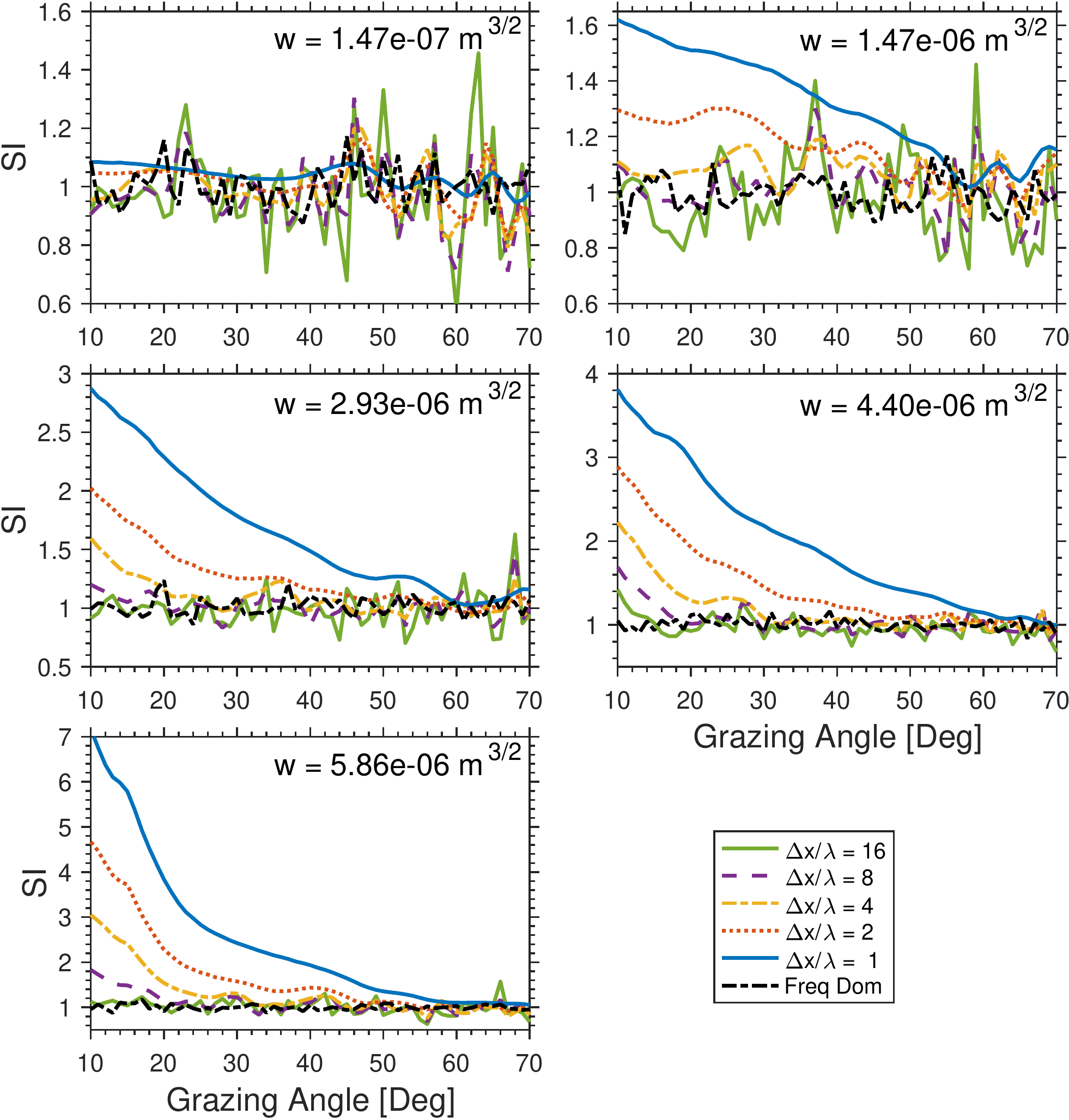}
  \caption{(color online) Scintillation Index for $f_0=$100 kHz and $\gamma=1.5$. Note that the vertical axis is different for each subfigure.}
  \label{fig:SI_100kHz_gamma1.5}
\end{figure}

\begin{figure}
  \centering
    \includegraphics[width=3.375in]{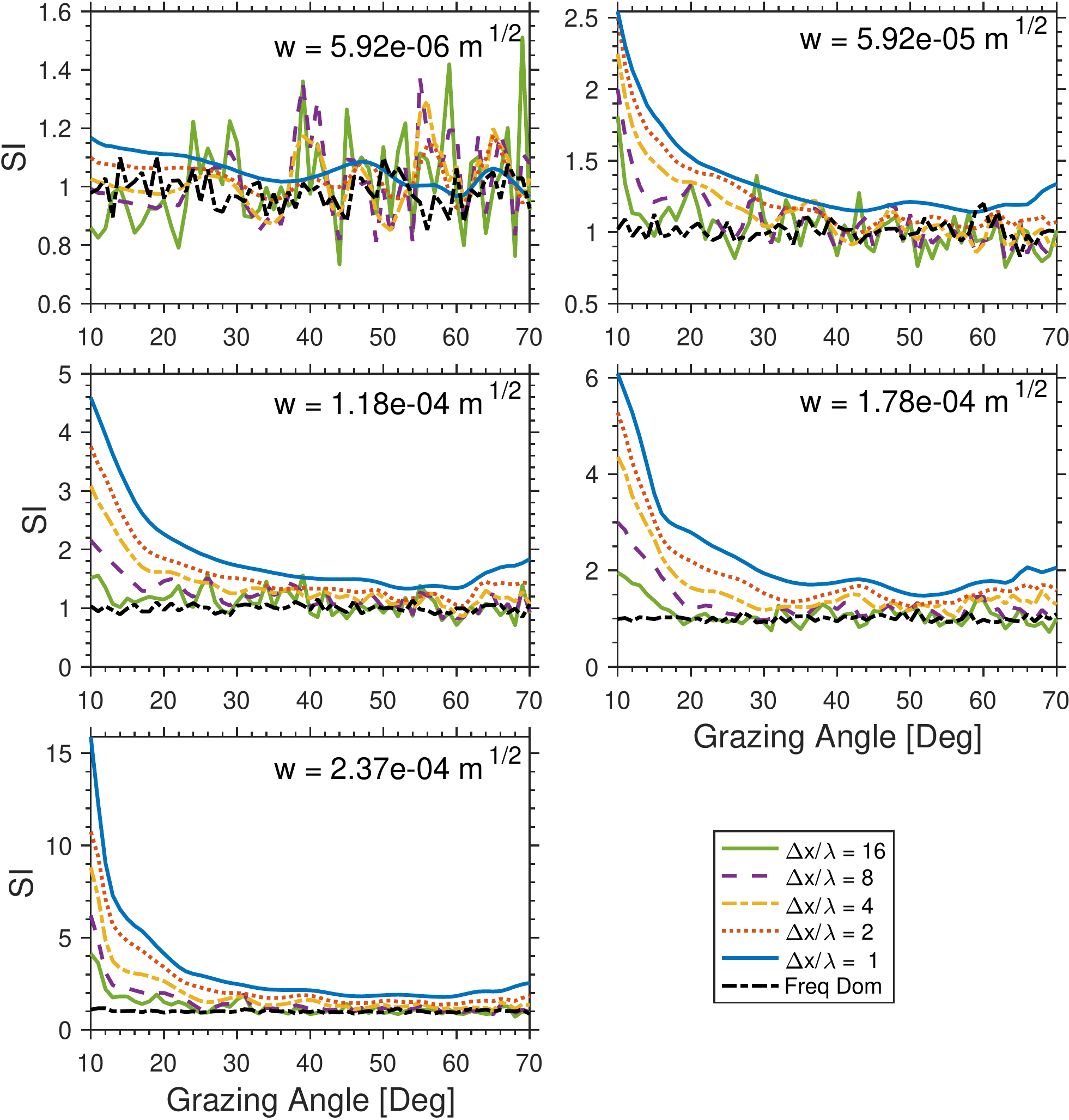}
  \caption{(color online) Scintillation index for $f_0$=100 kHz and $\gamma=2.5$. Note that the vertical axis is different for each subfigure.}
  \label{fig:SI_100kHz_gamma2.5}
\end{figure}

Results for backscattering strength as a function of grazing angle, $\theta_i$, resolution $\Delta X$, and spectral strength, $w$, are presented in Fig. \ref{fig:SS100kHz_gamma2} for $f_0 = 100$ kHz and $\gamma=2$. The numerical method can handle bistatic geometries (with differing $\theta_i$ and $\theta_s$), but only monostatic geometries are presented here. Broadband scattering strength, $\sigma_{bb}$ as well as the frequency domain version, $\sigma_f$ is plotted on the vertical axis, with grazing angle on the horizontal axis. The angles have a lower limit of 10$^\circ$ grazing angle due to finite surface length, and an upper limit of 70$^\circ$, due to the difficulty in estimating the broadband cross section near vertical using broadband pulses \cite{hefner_2015,hellequin_etal_2003}. Each resolution is plotted as a different color and line style, with the frequency domain version as a black dashed-dot line. Results for different spectral strengths are on the same figure since they are well separated from one another, with the smallest $w$ corresponding to the lowest scattering strength. For each value of the spectral strength, the frequency-domain scattering cross section is indistinguishable from the broadband version.

To compare these results more closely, the broadband cross section is divided by the frequency-domain cross section and the dB value taken. This quantity, which is called the scattering strength difference, is plotted in Fig. ~\ref{fig:SSError_100kHz_gamma2}. At the largest spectral strength, some systematic oscillations as a function of $\theta_i$ are present, but cannot be easily disentangled from the rapid Monte-Carlo fluctuations. Other than that case, all differences appear to be random. The standard deviation of the ratio $\sigma_{bb}/\sigma_f$ across angles between 10 and 70 degrees grazing is about 4\% for $\Delta X/\lambda_0=1$, and about 10\% for $\Delta X/\lambda_0=16$. Note that since this ratio is between two random variables, the uncertainty may be higher than the theoretical uncertainty for $\sigma_{bb}$ or $\sigma_f$ alone. These numbers are relatively consistent across all spectral strengths. Larger uncertainty for long pulses is a consequence of having fewer independent samples per roughness realization. The uncertainty for long pulse lengths is consistent with the theoretical uncertainty shown in Table~\ref{tab:numberOfIndependentSamples}, but the uncertainty for short pulse lengths is higher by about factor of 2.5. This increased uncertainty is likely a result of the departure from exponential intensity statistics for short pulses. Broadband scattering strength and the frequency domain scattering strength are indistinguishable, to within the Monte-Carlo error of the numerical simulations. This result confirms that the broadband scattering strength is a robust metric across different resolutions for scattering from a power-law seafloor with spectral exponent $\gamma=2$.

The frequency domain scattering cross section exhibits power-law frequency dependence, as seen by the nearly $f^1$ trend in Fig.~\ref{fig:dataProcessingSteps}(a). When incoherently averaged in the frequency domain, weighted by $S(f)$, the frequency dependence in the cross section for $\gamma \in [1,3]$, results in a negligible bandwidth dependence, under 4 percent. If $\sigma_f$ exhibited very strong frequency dependence, e.g. with sharp peaks seen in scattering from layered seafloors (e.g. \cite{jackson_ivakin_1998,Jackson2020}), then more significant bandwidth dependence may be observable.

The scintillation index for this case is plotted in Fig.~\ref{fig:SI_100kHz_gamma2} for broadband signals as well as the single frequency case. Each spectral strength is plotted in its own subfigure, and each resolution has its own line within the subfigure. The vertical axis is $SI$, and the horizontal axis is grazing angle in degrees. The narrowband result is approximately unity for the entire angular domain shown, for all spectral strengths. This is the expected result if the central-limit theorem is employed. For the broadband signals, there is a profound dependence on resolution, with the scintillation index increasing as the resolution cell becomes small. This behavior can be seen in the example realization shown in Fig.~\ref{fig:dataProcessingSteps}(c), in which the intensity peaks become higher as $\Delta X$ becomes small. Additionally, holding resolution constant, the scintillation index increases as the grazing angle becomes small, monotonically for this case. For most broadband cases, $SI$ asymptotically approaches unity as grazing increases to its upper limit.

However, for the highest resolution cases, $\Delta X/\lambda_0=( 1,2)$, and the largest spectral strengths, this high angle asymptote is greater than one, indicating that for all angles examined here, scattered complex pressure magnitude is non-Rayleigh. In \cite{lyons_etal_2016} a K distribution was required to describe the pdf of the scattered field at moderate grazing angles, which agrees with this result. The Monte-Carlo fluctuations are significantly less than the difference between the high-angle $SI$ asymptote and unity, indicating that this is a statistically significant finding.

\cite{lupien_1999} also observed non-Rayleigh scattering for broadband scattering from rough surfaces with a power-law exponent of $\gamma=3$, but statistical tests barely rejected the Rayleigh distribution. That analysis did not remove the effect of the Gaussian taper, so the conclusions are not comparable to the present work.

Broadband and single-frequency scattering strength and $SI$ were also computed for 10 kHz, $\gamma=2$, and spectral strengths that were ten times the value in the previous section. The relative resolution, $\Delta X/\lambda_0$ was held constant, but consequently the resolution $\Delta X$ was a factor of 10 larger. These parameters were chosen such that the dimensionless second-order quantities were the same as in the previous 100 kHz simulations. This scaling is only true for $\gamma=2$, and would be modified if $\gamma$ were a different value. It was found that the scattering strengths, scattering strength dB error, and the scintillation index were the same for the 10 kHz and 100 kHz cases, to within Monte-Carlo fluctuation. This set of simulations was performed to verify that characterizing the simulations non-dimensionally was valid. Since plots for the 10 kHz case do not add significantly new information, they are not shown here. These results indicate that departure from Rayleigh statistics is not isolated to very high-frequency imaging systems, and may occur in lower-frequency sonar systems as well, so long as the seafloor has the appropriate roughness parameters.

The spectral exponent was changed to $\gamma=1.5$ to examine the effect of changing the shape of the power spectrum. New values of $w$ were used, as specified in Table~\ref{tab:roughnessParameters}. Again, $\sigma_{bb}$ and $\sigma_f$ are the same. Scattering strength comparisons and the scattering strength ratio are not shown. The scintillation index is plotted in Fig.~\ref{fig:SI_100kHz_gamma1.5}, and is seen to depend on angle, resolution, and spectral strength, as in the previous case.  The rough interfaces have the same effective large-scale slope for $\gamma=2$ and $\gamma=1.5$ at $\Delta X/\lambda_0=1$. $SI$ qualitatively depends similarly on spectral strength and resolution, as compared to the $\gamma=2$ case. Quantitatively, the values of $SI$ are slightly different than the $\gamma=2$ case, but are similar.

Finally, the spectral exponent was changed to $\gamma=2.5$. The values of spectral strength can be found in Table~\ref{tab:roughnessParameters}. Again, the broadband scattering cross section, and the frequency-domain version computed at the center-frequency were the same to within the Monte-Carlo error of the simulations, and are not shown here. The scintillation index is plotted in Fig.~\ref{fig:SI_100kHz_gamma2.5}, and again depends on resolution, spectral strength and grazing angle. As the spectral strength is increased in the same proportions as the earlier plots (the second through fifth spectral strengths are 10, 20, 30, and 40 times the smallest spectral strength respectively), the scintillation index increases much more rapidly than either the $\gamma=2$ and $\gamma=1.5$ cases. For the four largest values of spectral strength, the $SI$ is elevated at the higher angles as well. The smallest resolution cases also have elevated $SI$ for the entire angular domain, and the four highest spectral strength values.

\section{Discussion}
\label{sec:discussion}
In the results presented in Sec.~\ref{sec:results}, it was shown that the broadband scattering strength is indistinguishable from frequency-domain scattering strength, and is independent of pulse length. This conclusion is not surprising. However, as the pulse length changes, the properties of the ensemble used to estimate scattering strength changes as well. It is encouraging to see that although the ensemble is changing with respect to resolution (i.e the rough patch within a resolution cell is different for each resolution), $\sigma_{bb}$ is invariant to pulse length. It is expected that this result holds for 3D environments as well if the roughness is isotropic, although numerical tests using this geometry are needed. Based on these results, high-resolution systems should be able to reliably estimate scattering strength, and this work also confirms that it may be a stable quantity to use for seafloor remote sensing. However, for highly anisotropic, non-stationary scenarios, such as those studied by \cite{olson_etal_2016,olson_etal_2019,lyons_etal_2010}, the measured scattering strength may depend on pulse length.

The scintillation index (also called structure \cite{wang_bovik_2002}, lacunarity \cite{williams_2015}, or contrast \cite{marston_plotnick_2015}) was shown to be is highly dependent on all the parameters studied: resolution, grazing angle, spectral strength, and spectral exponent. For moderate to low grazing angles, $SI$  monotonically increases as grazing angle decreases, resolution cell decreases, and spectral strength increases. SI is similar for $\gamma=1.5$ and $\gamma=2$, but is larger for $\gamma=2.5$, for the values of $w$ chosen for these numerical experiments. Contrary to scattering strength, $SI$, and therefore the scattering process in general, is fundamentally different in the frequency and time domains for broadband pulses.

In \cite{lyons_etal_2016}, it was hypothesized that the physical cause of heavy-tailed statistics in high resolution sonar imagery was local tilting of the seafloor due to roughness wavelengths larger than the acoustic resolution. The scattered pressure amplitude (envelope) was modeled as a product between a random variable due to sub-resolution roughness, and a random variable that took into account the effect of tilting by longer wavelengths  -- the small-scale scattering strength evaluated at the nominal grazing angle modified by the local slope. Local tilting (due to large-wavelength roughness components) modulates the Rayleigh-distributed field (due to small-wavelength components) and causes the $SI$ to be greater than unity. An rms slope with an upper cutoff related to the acoustic resolution was used as the input parameter to a simplified version of the composite roughness model \cite{mcdaniel_gorman_1983}, which was then used to compute the scintillation index.

Here, the role of slopes at scales at or larger than the pulse length on the intensity fluctuations is investigated. A representative function is used to model the effect of slope modulation that is inspired by the composite roughness model. The composite roughness approximation \cite{mcdaniel_gorman_1983} uses separate scattering models on the small and large-scale surfaces. The validity of the composite model is subject to the validity of the perturbation approximation applied to the small-scale surface, and the Kirchhoff approximation applied to the large-scale surface. The perturbation approximation is valid when $k_0 h_s$ is small \cite{thorsos_jackson_1989}, where $h_s$ is the rms roughness of the small-scale roughness. This parameter can be found by integrating the roughness spectrum over wavenumbers with magnitudes between $\pi/(A\Delta X)$ and $\pi/\ell_0$. In this work, $k_0 h_s$ was less than 2.0 for all simulations. Since perturbation theory must be applicable on the small-scale surface, $k_0 h_s <1$ was required. This is possible if the highest spectral strength for each value of $\gamma$ is ignored, and the restriction $\Delta X/\lambda_0 \leq 8$ is made. We found that scattering strength was well modeled by perturbation theory for a few cases where $k_0 h >1$ (including all the wavenumber components), but exclude these cases since the validity region of this model for general power-law roughness spectra has nost been previously studied.

In \cite{mcdaniel_gorman_1983, jackson_composite}, the validity of the Kirchhoff approximation was stated to be a function of the rms radius of curvature of the large-scale surface. However, \cite{thorsos_1988}, found that the radius of curvature is not important for the Kirchhoff approximation, but that the characteristic length of the surface, $L_c$, must be large compared to the wavelength. In the separation of scales performed here, the characteristic length of the large scale surface is approximately equal to $\Delta X$, which is always $\lambda_0$ or greater. Therefore $k_0 L_c\geq2\pi$, which is typically acceptable for the Kirchhoff approximation \cite{thorsos_1988}. At low grazing angles, which are most relevant here, the Kirchhoff approximation fails due to the presence of multiple scattering, which is important when $\theta_i$ and $\theta_s$ are each less than or equal to $2\epsilon$, twice the large-scale rms slope angle \cite{thorsos_1988,thorsos_jackson_1991}. Shadowing is another source of inaccuracy, which occurs when $\theta_i$ and $\theta_s$ are less than or equal to $\epsilon$ \cite{thorsos_1988}. The large-scale slope angles can be seen in Table~\ref{tab:roughnessParameters}. For the rms slope angles that are based on the acoustic resolution as the upper cutoff, the global maximum is about 5$^\circ$. Since this number is comparable to the smallest grazing angles examined here, both multiple scattering and shadowing may be present in these cases.

To explore the local tilting hypothesis, the relationship between intensity as a function of space $I_{\Delta X}(x)$, by which is $I(t^\prime)$ for a given resolution mapped to $x$ through $x=-ct^\prime/(2\cos\theta)$ for backscattering, with the large-scale surface slope at position $x$ is investigated. $I_{\Delta X}(x)$ should decrease if the large-scale slope at $x$ is positive, and increase if negative. Consequently there should be a statistical correlation between $I_{\Delta X}(x)$, the intensity at a given resolution at position $x$, and $-s_{\Delta X}(x)$, the negative of the slope field low-pass filtered to remove wavelengths shorter than $\Delta X$. More specifically, the scattered intensity from the integral equations can be compared to the intensity produced by the effect of tilting in the composite roughness approximation. A simplified version of this approximation is used, which is explained in the following paragraphs.

The composite model requires two types of averages, one over the small-scale and one over the large-scale. The large-scale average is a simple average of the small-scale scattering cross section over the pdf of the large-scale slopes. The small scale average results in the scattering cross section from the small-scale roughness. Using the results of \cite{mcdaniel_gorman_1983}, the composite model may be written as (after converting between differing conventions of the roughness power spectrum, and adapting to 1D roughness used here),
\begin{align}
	\sigma_{CR} &= 4 k^3 \langle \sin(\theta_i - \epsilon_L)^4  W_s(\Delta K_{mod})\rangle_{L} \\
	\Delta K_{mod} &= \Delta K + \frac{df_L}{dx}\Delta k_z
	\label{eq:composite_}
\end{align}
where $\langle\cdot \rangle_L$ denotes averaging over the large-scale slope, and $\epsilon_L=\tan^{-1}(df_L/dx)$ is the large-scale slope angle at position $x$. The power spectrum of the small-scale roughness is denoted $W_s(\Delta K_{mod})$, where the modified Bragg wavenumber, $\Delta K_{mod}$ is specialized to the backscattering case and takes into account the effect of tilting on the Bragg components. The difference between the vertical component of the scattered and incident vertical wavenumber is $\Delta k_z = 2k_0 \sin\theta_i$, and the horizontal wavenumber difference is $\Delta K = 2k_0 \cos\theta_i$. Although not present in most applications of the composite roughness model (e.g. \cite{Kuryanov1963,Bachmann1973,mcdaniel_gorman_1983}), the modulation of the Bragg wavenumber due to large-scale slopes may be significant. The form of the Bragg modulations is given in an intermediate result (Eq. (22)) of \cite{mcdaniel_gorman_1983}, and adapted to the present notation. This model was formulated in the frequency domain, and is not directly applicable to these broadband simulations. However, in the broadband case, the effect of tilting would be preserved, even if the scattered power from the small-scale surface is altered.


Because of this simplification and the broadband nature of the simulations, this model is called the slope modulation model (SMM) instead of the two-scale, or composite roughness model. Terms that are constant as a function of the large-scale slope were ignored in the SMM. Averaging over the large-scale slopes was not performed in the model, to compare fluctuations produced by tilting to the fluctuations from the integral equation results for an given ensemble. The intensity fluctuations caused by tilting can be written as,
\begin{align}
  \begin{split}
I^{SMM}_{{\Delta X}}(x) = \sin^4&\left( \theta_i - \tan^{-1}\left[\frac{df_{\Delta X}}{dx}(x)\right]\right) \\%
						  \times&W(\Delta K + \Delta k_z\,\frac{df_{\Delta X}}{dx}(x))
\end{split}\label{eq:modified_composite_roughness}
\end{align}
where the subscript $\Delta X$ on the intensity and slope explicitly denotes that a Gaussian function with 3 dB width $\Delta X$ was used to filter the large-scale slopes. The filtered slope is defined by
\begin{align}
	\frac{df_{\Delta X_{surf}}}{dx}(x) = \int\limits_{-\infty}^{\infty}\frac{df}{dx}(x^\prime - x) e^{-x^{\prime 2} \ln(4)/(\Delta X_{surf})^2}\mathrm{d} x^\prime\, .
	\label{eq:surfaceFilter}
\end{align}
Eq.~(\ref{eq:modified_composite_roughness}) has been averaged over the small-scale roughness, but not the large-scale slopes. It is important to note that in the integral equation simulations, it is impossible to perform this partial averaging. Thus, taking the variance of Eq.~(\ref{eq:modified_composite_roughness}) does not result in the scintillation index of the scattered pressure, but only the variance of slope-induced fluctuations. Taking the limit of zero large-scale rms slope in the previous equation results in no slope-induced fluctuations, and fluctuations in the total scattered field would have a Rayleigh distribution, with a scintillation index of unity. It is strongly emphasized that this model is only a representative function of the effect of slopes at the scale of the pulse resolution, and is not an adequate model for the scattered intensity. This inadequacy is due to the broadband nature of these simulations, and the formulation of the composite model in the frequency domain. The integral equation results are considered to be the accurate ``ground truth'' here.

Although not shown in (\ref{eq:modified_composite_roughness}), $I^{SMM}_{{\Delta X}}(x)=0$ if the argument to the sine function is less than zero, to include a rudimentary form of shadowing. This form of shadowing has a small effect, and the results presented below are essentially unchanged if this simple form of shadowing is left out. Non-local shadowing may have a significant effect, but is not examined here. The simplifications used here are employed because a rigorous formulation of the composite roughness approximation for very broadband signals is not available in the literature. Development of such a model, and comparison to results using the broadband integral equation technique developed here are both fruitful areas for future work.

Similarity between the intensity from the numerical solutions and model predictions can be quantified in a crude but straightforward manner using the Pearson product-moment correlation coefficient, $\rho$, defined for random variables $U$ and $V$.
\begin{align}
	\rho \left( U,V \right) = \frac{\langle \left( U - \langle U \rangle \right)  \left( V - \langle V \rangle \right) \rangle }{\sqrt{ \langle \left( U -  \langle U \rangle \right)^2\rangle  \langle \left( V - \langle V \rangle \right)^2 \rangle }}
\end{align}
This coefficient quantifies the linear variation between a dependent and an independent variable. The correlation coefficient is used here because a rigorous application of the composite roughness approximation was not used, and absolute intensity cannot be compared directly.

The local tilting hypothesis is tested by forming the correlation coefficient between the scattered intensity from numerical simulations, and $I^{SMM}_{{\Delta X}}(X)$. Both the acoustic resolution and the surface filter size are varied. The surface filter scale that maximizes the correlation for each acoustic resolution is estimated, and this process is repeated for all acoustic resolutions. The acoustic resolution is $\Delta X$, and the surface filter size is $\Delta X_{\textrm{surf}}$. The $\Delta X_{\textrm{surf}}$ that maximizes $\rho$ is denoted $\Delta X_{\textrm{max}}$. If $\Delta X_{\textrm{max}}$ varies in proportion to $\Delta X$, then it may be concluded 1) that slope modulation is responsible in part for the intensity fluctuations, and 2) that slopes at (or larger than) the scale of the acoustic resolution are responsible in part for the fluctuations. Note that the surface filters used here are zero-phase, acausal filters (Eq.~(\ref{eq:surfaceFilter})), and are applied in the same way as the Fourier synthesis used to obtain the time-domain scattered pressure, Eq.~(\ref{eq:fourierSynthesis}).

\begin{figure}
  \centering \includegraphics[width=3.375in]{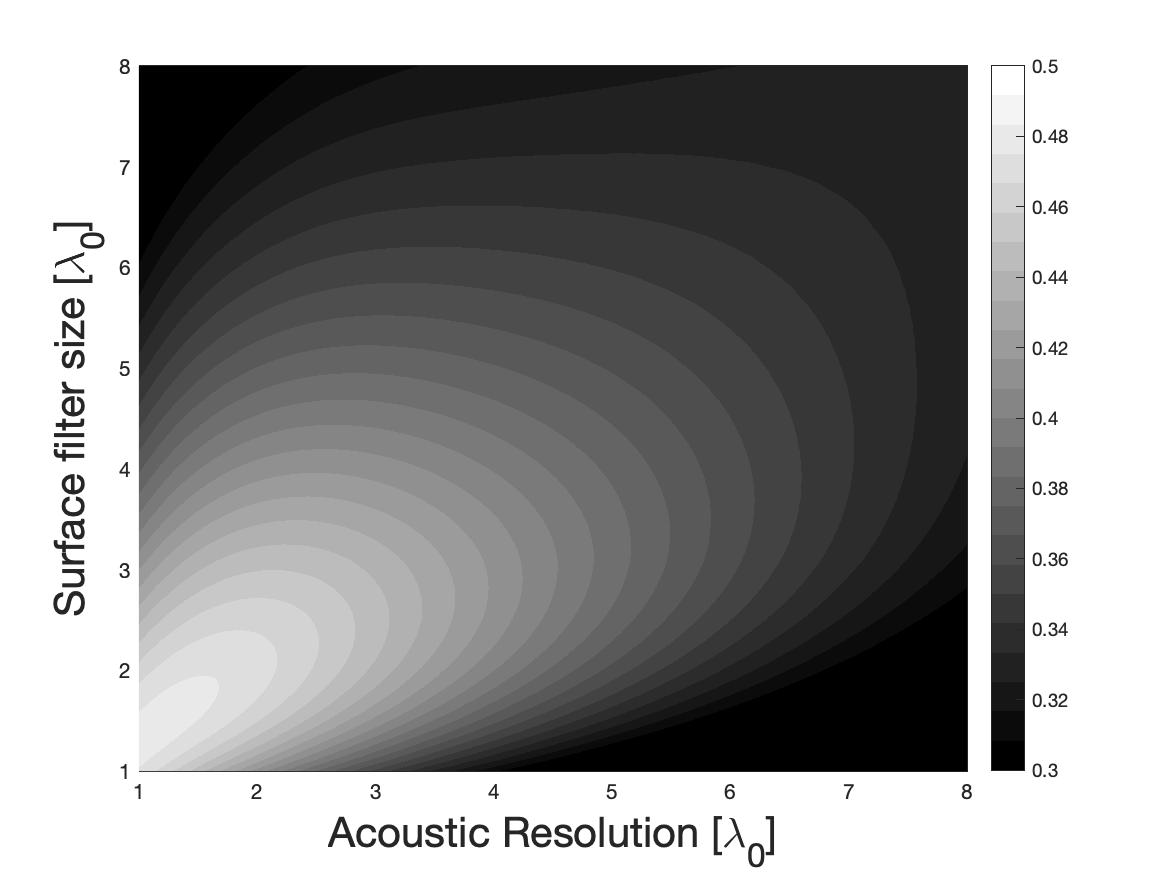}
  \caption{Correlation coefficient as a function of the acoustic resolution, $\Delta X$, and the surface filter size, $\Delta X_{surf}$. The Gaussian pulse shape has been used as the surface filter here, as specified by Eq.~\ref{eq:surfaceFilter}.}
  \label{fig:correlationCoeff}
\end{figure}

\begin{figure}
  \centering \includegraphics[width=3.375in]{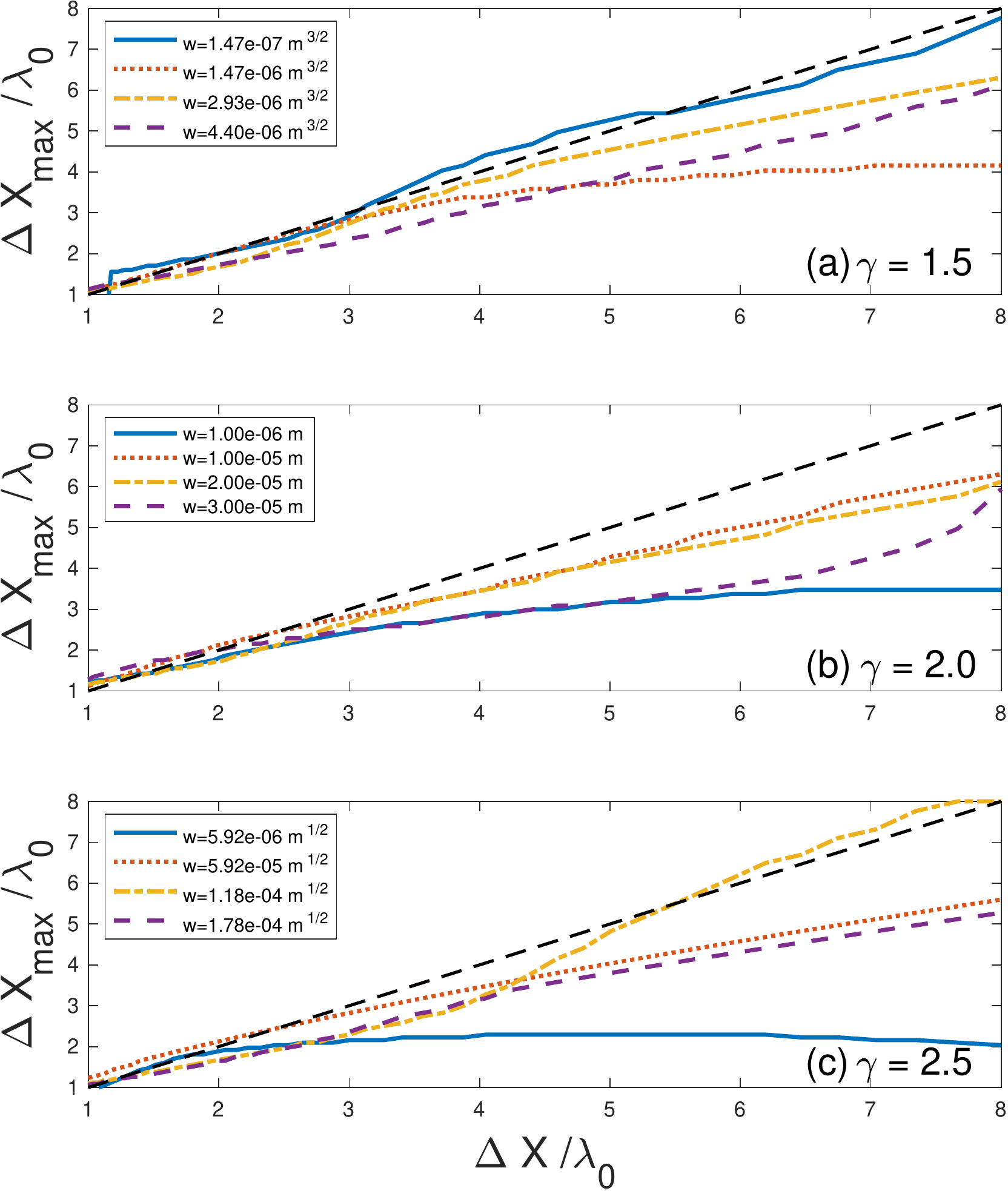}
  \caption{(color online) Surface slope filter scale that maximizes the product moment cross correlation coefficient for each acoustic resolution. Grazing angle has been held constant at $20^\circ$, and each line represents a different spectral strength. Each subplot contains a different spectral exponent with (a)$\gamma=1.5$, (b)$\gamma=2$, and (c),$\gamma=2.5$. Dashed black lines have a slope of unity and intercept of zero for reference. Note that the Gaussian pulse shape, Eq.~(\ref{eq:surfaceFilter}) has been used as the surface filter.}
  \label{fig:maxCorrelationScale}
\end{figure}

The correlation coefficient for parameters $\gamma=2$, $\theta_i=20^\circ$, and $w=3\times10^{-5}$m is plotted in Fig.~\ref{fig:correlationCoeff}. $\Delta X$ is on the horizontal axis, $\Delta X_{\textrm{surf}}$ is on the vertical axis, and $\rho$ is denoted by grayscale. Holding $\Delta X$ constant, there is a distinct peak in $\rho$ as a function of $\Delta X_{\textrm{surf}}$. The peak value of $\rho$ increases as the acoustic resolution becomes small, indicating that the role of slope modulation is greater for smaller resolution cells. Additionally, the peak location in $\Delta X_{\textrm{surf}}$ varies with $\Delta X$, indicating the local tilting hypothesis may be correct. This plot has a similar structure for other $\theta_i$, $\gamma$, and $w$.

$\Delta X_{\textrm{max}}$ as a function of $\Delta X$ is plotted in Fig.~\ref{fig:maxCorrelationScale}. $\theta_i$ is constant at $20^\circ$ grazing angle, and each spectral strength is plotted as its own line. The different values of $\gamma$ appear in subfigures. For each spectral strength and exponent, $\Delta X_{\textrm{surf}}$ varies monotonically with the acoustic resolution, except for the smallest spectral strength for $\gamma=2.5$. A line with unit slope and zero intercept is also plotted for reference. For many cases, the slope of these lines is approximately unity when the acoustic resolution is small. Some of the lines retain that slope when the acoustic resolution is larger, but others taper off and have a smaller slope. For example the smallest roughness cases in $\gamma=2$ and $\gamma=2.5$ curve downards to an approximately constant function at large $\Delta X/\lambda_0$. For these two curves in particular, the scintillation index was approximately unity, and therefore slope modulation was not required to explain their behavior. Other cases where the slope of these lines becomes small for large $\Delta X / \lambda_0$, but result in in $SI>1$, indicate that slopes at scales smaller than the pulse resolution are the most important. Future work modeling broadband scattering from power law surfaces is needed to explain this behavior.


The monotonic dependence of $\Delta X_{\textrm{max}}$ on $\Delta X$ indicates that the large-scale slope near the acoustic resolution accounts for a significant part of the intensity fluctuations for short pulses. The unit slope for some of these cases indicates that the 3 dB width of the point spread function is sometimes the appropriate scale for calculating the large-scale slope, and is a good estimate for all cases. For longer pulses, the lines in Fig.~\ref{fig:maxCorrelationScale} have slopes generally less than unity, meaning that the horizontal roughness components that are important for slope modulation are not exactly the same as the 3 dB width of the incident intensity, but are proportionally slightly smaller. This departure from unit slope may indicate that the slope modulation model is not sufficient to explain the data, and other effects, such as multiple scattering or shadowing may be present. The same analysis was performed for the large-scale surface heights, and no systematic trends were found. Therefore, the large-scale rough interface at a single point is uncorrelated with the scattered intensity mapped to that point.

\begin{figure}
  \centering
    \includegraphics[width=3.375in]{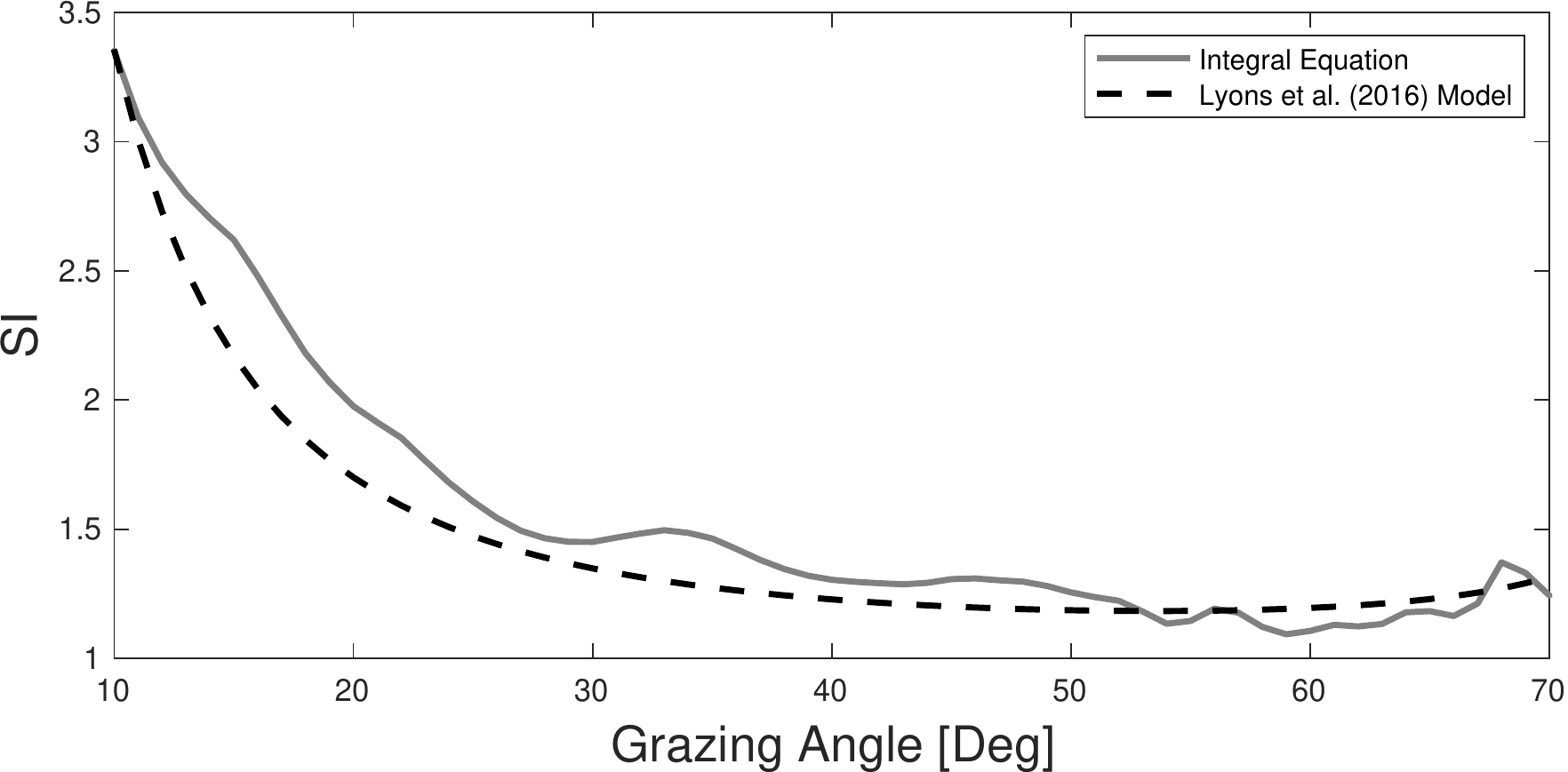}
  \caption{Comparison of the $SI$ from the integral equation results with the model of \cite{lyons_etal_2016}, for the parameters $w=3\times10^{-5}$m, $\gamma=2$, and $\Delta X=2\lambda_0$.}
  \label{fig:SI_IE_model_comparison}
\end{figure}

To understand what kind of intensity fluctuations can be caused by this model when for the large-scale rms slope values in these numerical simulation, the model for $SI$ from \cite{lyons_etal_2016} was applied to the case of $w=3\times10^{-5}$ m, $\gamma=2$, and $\Delta X=2\lambda_0$. This model accounts for fluctuations due to large-scale tilting, as well as Rayleigh-distributed scattering from the small-scale roughness to produce the scintillation index. In the model of \cite{lyons_etal_2016}, $SI>1$ if the large-scale slope variance is greater than 0, and the small-scale scattering cross section is neither constant nor linear in $\theta_i$. The rms slope was taken from Table~\ref{tab:roughnessParameters}, and was $2.99^\circ \approx3^\circ$ ($s_{2A\lambda_0}$ in Table~\ref{tab:roughnessParameters}). A model-data comparison is shown in Fig.~\ref{fig:SI_IE_model_comparison}. The model follows the same trends as the integral equations. It is slightly higher than 1 at high grazing angles, and increases as the grazing angle becomes small. However, the model $SI$ increases less rapidly than the integral equation simulation at low grazing angles. Given the evidence from Figs.~\ref{fig:correlationCoeff}, and \ref{fig:maxCorrelationScale}, as well as evidence that the composite roughness model produces non-Rayleigh behavior that is similar to the integral equations it may be concluded that the tilting hypothesis is responsible in part for the intensity fluctuations. Note that both the $\sin^4$ term and Bragg modulation term from Eq.~(\ref{eq:modified_composite_roughness}) were included to account for fluctuations due to slopes. Neglecting the Bragg term decreases $SI$ by about 5\% at 10$^\circ$, and about 20\% at 70$^\circ$ grazing angle. Including modifications to the Bragg wavenumber is responsible for the good agreement at high and moderate grazing angles.

Since the model of \cite{lyons_etal_2016} underestimates $SI$, especially for small grazing angles, the tilting hypothesis cannot be fully responsible for heavy-tailed scattering seen here. We suspect multiple scattering and shadowing may be responsible for the model-simulation mismatch at low angles, since the large-scale rms slope is 3$^\circ$ for the surface in Fig.~\ref{fig:SI_IE_model_comparison}, and is comparable to the smaller range of grazing angles \cite{thorsos_1988,thorsos_jackson_1991}. Shadowing may increase $SI$, since it increases the degree of intensity fluctuations. Multiple scattering may move $SI$ towards unity, since scattering from multiple locations on the rough interface may increase the effective ensonified length, and contribute an additive Gaussian component to the scattered pressure.

A physically accurate theoretical model that includes all of these effects is evidently required to predict the scintillation index at low grazing angles. Such a model is, at this time, not available, and is a fruitful opportunity for future research. The independence of the broadband scattering cross section on resolution established here imposes a useful constraint. Any theoretical model that breaks apart the solution of the exact integral equation into physically interpretable phenomena (such as tilting, shadowing, or multiple scattering) must also satisfy this constraint.

\section{Conclusion}
\label{sec:conclusion}
In this work Fourier synthesis combined with numerical solution of the Helmholtz integral equation was used to analyze the scattered field in terms of the broadband scattering cross section and scintillation index. The dependence of these two quantities on acoustic resolution was examined to understand the effects that contemporary high-resolution acoustic imaging systems have on quantitative measurements of seafloor scattering. For power-law surfaces, the scattering strength is independent of pulse length, which indicates it is a stable quantity to use across measurement systems with different geometries (at least for the kinds of rough surfaces examined here). The scintillation index depends strongly on pulse length, which indicates that the scattering processes is fundamentally different in the time and frequency domains, and that further research is needed to understand or predict intensity fluctuations in high resolution broadband sonar systems.

Although simulations were performed in two dimensions, these results may be present in three dimensions as well, if the roughness spectrum is assumed to be isotropic. The exact values of the scintillation index will be different for 3D environments, as the rms slope is calculated differently, and out of plane effects may be important.

Heavy-tailed, or non-Rayleigh scattering is commonly observed in scattering measurements and is usually attributed to non-stationary, or patchy environments \cite{abraham_lyons_2004,lyons_etal_2009}. Heavy-tailed statistics have been observed in seemingly homogeneous seafloors by \cite{lyons_etal_2016}, and these numerical simulations have verified that statistically homogeneous surfaces can produce heavy-tailed statistics when interrogated by a broadband high-resolution system. The slope modulation model used in \cite{lyons_etal_2016} was investigated, and found that it is in part responsible for intensity fluctuations. Other sources of fluctuations, such as shadowing and multiple scattering, were found to be necessary to accurately model the scintillation index. Further research is required to better understand this aspect of seafloor scattering.

A few consequences are noted for heavy-tailed statistics arising in high-resolution systems in homogeneous roughness environments. Heavy-tailed statistics are a significant source of false alarms in acoustic target detection systems. Since scintillation index increases at low angles, long range systems may suffer from decreased performance. The benefit of high resolution systems, more pixels per target, may be offset due to the increased false alarms.

There may be some benefits of resolution dependence of the pdf of the scattered field. Some autofocus algorithms for synthetic aperture systems (e.g. \cite{blacknell_etal_1992,callow_thesis,marston_plotnick_2015}), use the scintillation index, or contrast as their cost function. If the acoustic field is entirely due to point scatterers, as is commonly assumed \cite{brown_etal_2017}, $SI$ will be unity for all resolutions. If an autofocus algorithm is applied, and the point spread function of the imaging algorithm becomes smaller, then the field will still be Rayleigh and contrast will not increase. Therefore, an autofocus algorithm based on $SI$ or lacunarity will not be sensitive to the focus, unless there are discrete scatterers in the scene. However, as shown here, $SI$ is a strong function of resolution for a statistically homogeneous power law surfaces, especially at low grazing angles. Improving the focus at low grazing angles will lead to an increase in $SI$, and thus the autofocus algorithm will be more sensitive to the actual degree of focus for seemingly featureless seafloors.

Median filters are often used to ``remove'' the speckle or intensity fluctuations from acoustic or electromagnetic images before use in remote sensing or target detection algorithms (e.g. \cite{Kwon2005,williams_2015,williams_2018,Galusha2018} and references therein). Although the broadband scattering cross section, which uses the arithmetic mean of the intensity, is insensitive to resolution, the median is not, since the median is highly dependent on the probability density function. Thus either the broadband cross section can be used as a resolution independent quantity that has a large variance, or its variance can be reduced with the consequence that the pixel intensity will no longer be directly related to broadband scattering strength.

\acknowledgments
The authors thank the US Office of Naval Research for financial support of this work through grants N00014-13-1-0056, N00014-18-WX-00769, N00014-19-WX-00427, and N00014-16-1-2268, as well as the Research Initiation Program of the Naval Postgraduate School. Numerical simulations were performed on the Hamming high-performance computing cluster at Naval Postgraduate School. The authors thank the anonymous reviewers for their comments on previous versions of the manuscript.

\appendix
\section{}
\label{sec:randomRoughSurfaceGeneration}
In this appendix, the steps are given to generate random realizations of a given roughness power spectrum. Here, the method of \cite{thorsos_1988} is followed, specifically Eqs. (17-18). The spatial frequency is $u = K/(2\pi)$. The sampling of this vector is set by the surface length, $L$, and sampling interval, $\delta x$. The number of points, $N$, is $L/\delta x$, and the spatial frequency spacing is $\delta u = 1/L$. A spatial frequency vector is specified, $u_m = m \delta u$ for $m \in [-N/2+1 , N/2]$. Then, a sampled power spectrum vector is created, $W_m = W(2\pi u_m)$. To create a random realization from this power spectrum, a sampled version of a randomized complex amplitude spectrum, $F$, is defined by
\begin{align}
	F_m = \sqrt{2 \pi W_j L} \mu_m
\end{align}
where $\mu_m$ are elements of a complex random vector with unit variance. To generate a real vector of heights, $f_n$, the amplitude spectrum $F$ must have conjugate symmetry. This property is achieved by setting
\begin{align}
	\mu_m = %
	\begin{cases}
	\frac{N(0,1)+ i N(0,1)}{\sqrt{2}} & m \in [1,N/2-1]\\
	\mu_{-m}^\ast & m < 0 \\
	N(0,1) & m=0,N/2
	\end{cases}\, ,
\end{align}
where $N(0,1)$ is an independent random draw from the normal distribution with zero mean, and unit variance. The first case creates random vector for positive spatial frequencies, the second case enforces conjugate symmetry. The third case stipulates that points in the sampled spatial frequency vector that do not have symmetric pairs (the point at zero frequency, and the single Nyquist point) are real. This method is valid for even values of $N$. For odd $N$, $m\in[-(N-1)/2 , (N-1)/2]$, and the lower Nyquist point, $m=-(N-1)/2$ must be the complex conjugate of the point $M=(N-1)/2$, and only the case $m=0$ should be real.

The rough interface $f(x)$ is found through the inverse discrete Fourier transform. For even $N$, this is
\begin{align}
	f_n = f(n\delta x) = \delta u \sum\limits_{m = -N/2 +1}^{N/2}F_m e^{i 2\pi \frac{m n}{N} }\, .
	\label{eq:roughSurfaceSynthesis}
\end{align}
For odd $N$, the sum is the same, except that the limits are from $m=-(N-1)/2$ to $(N-1)/2$. In practice, this sum is approximated using the fast Fourier transform algorithm. The MATLAB computer language was used to generate the rough surfaces, which itself uses the FFTW library (fastest Fourier transform in the west) \cite{Frigo2005}. This library requires that the input spatial frequency spectrum be contained in a vector starting with the $m=0$ point, followed by the points with $m>0$, followed by the points with $m<0$. The FFTW library also divides by $N$ when performing the inverse FFT, so the output must be multiplied by $N$, and by $\delta u$ to form the sum in Eq.~(\ref{eq:roughSurfaceSynthesis}).

\section{}
\label{sec:energyFluxAppendix}
Here, the effective energy flux is calculated for the incident field used in this work. Effects of the $w_t$ term in the extended Gaussian beam are not included, since analytically tractable results are not available for that case. Although the full extended Gaussian beam was used as the incident field in the numerical simulations, the error introduced by omitting it in calculating the energy flux was insignificant, as the parameter $(k_0 g \sin\theta_i)^{-1}$ was always small, with a maximum of 3.7$\times10^{-3}$.

Setting $w_t = 0$, the incident pressure for a broadband, tapered pulse is given by
\begin{align}
	\begin{split}
	p_i(\mathbf{r},t) = p_0 &\psi(\mathbf{r})\\
	\times & \exp\left(i\omega_0 (t - t_i(\mathbf{r})\right)\\
	\times &\exp\left( - (t - t_i(\mathbf{r}))^2/\tau^2\right)\, ,
	\end{split}
	\label{eq:modIncidentPressure}
\end{align}
where
\begin{align}
	t_i(\mathbf{r}) =\frac{\mathbf{r}\cdot \hat{\mathbf{k}}_i}{c}= \frac{-x\cos\theta_i - z\sin\theta_i}{c},
\end{align}
and
\begin{align}
	\psi(\mathbf{r})= \exp\left(-(x - z\cot\theta_i)^2/g^2 \right)
\end{align}
is the spatial taper. The vertical component of the acoustic particle velocity can be computed as \cite[p. 19]{pierce_acoustics}
\begin{align}
	v_z(\mathbf{r},t) &= \frac{\partial}{\partial z}\Phi(\mathbf{r},t), \label{eq:vel_as_potential}
\end{align}
and the acoustic pressure as
\begin{align}
	p(\mathbf{r},t) &= -\rho_0 \frac{\partial}{\partial t} \Phi(\mathbf{r},t), \label{eq:presssure_as_potential}
\end{align}
where $\Phi$ is the acoustic velocity potential. Using Eq.~(\ref{eq:modIncidentPressure}) in (\ref{eq:presssure_as_potential}), the velocity potential can be computed using Eqs.~3.322(1-2) from \cite{grad_ryz},
\begin{align}
	\begin{split}
	\Phi(\textbf{r},t) &= \frac{-p_0}{\rho_0}
\psi(\mathbf{r}) M(t-t_i(\mathbf{r}),f_0,\tau)	\end{split}
	\label{eq:velocityPotential}
\end{align}
where
\begin{align}
	M(t,f_0,\tau) =\frac{\sqrt{\pi}\tau}{2} e^{-\pi^2 f_0^2 \tau^2}\erf\left[t/\tau -i\pi f_0 \tau\right],
\end{align}
and $\erf\left[z\right]$ is the error function \cite[p. 297]{abramowitz}.
If the opposite time convention is used, then the minus sign inside the error function argument must be changed to a plus. In the integration used to compute Eq.~(\ref{eq:velocityPotential}), an integration constant independent of time is required, but is assumed to be zero here, since derivatives are taken to compute $p$ and $v_{z}$.

Inserting Eq.~(\ref{eq:velocityPotential}) into Eq.~(\ref{eq:vel_as_potential}), one obtains
\begin{align}
	\begin{split}
	v_{iz}(\mathrm{r},t&) = -\frac{p_0}{\rho_0}\left[\frac{2\cot\theta_i}{g^2}(x - z\cot\theta_i)\psi(\mathbf{r})M(t-t_i(\mathbf{r}),f_0,\tau)\right. \\
	&\left.+\frac{\sin\theta_i \psi(\mathbf{r})}{c}\exp\left(i\omega_0 (t - t_i(\mathbf{r})) -(t - t_i(\mathbf{r}))^2/\tau^2  \right)\right]
	\end{split}
	\label{eq:incidentVerticalVelocity}
\end{align}

To compute the effective energy flux passing through the $z=0$ plane, Eq.~(\ref{eq:modifiedEnergyFlux}) is computed using Eqs.~(\ref{eq:modIncidentPressure}) and (\ref{eq:incidentVerticalVelocity}). The time variable is set to $t = R/c+t^\prime$, to match the convention in Eq.~(\ref{eq:fourierSynthesis}). Making these substitutions,
\begin{align}
	E_f^\prime =\frac{|p_0|^2}{2\rho_0}\left(\frac{2\cot\theta_i}{g^2}I_1 + \frac{\sin\theta_i}{c}I_2 \right)\, .
	\label{eq:energyflux_I1_I2}
\end{align}
The two integrals are defined as
\begin{align}
	\begin{split}
	I_{1} &=\frac{\sqrt{\pi}\tau}{2}e^{-(\omega_0\tau/2)^2} \\
		& \times Re\biggl[\int\limits_{-\infty}^{\infty} x\, \mathrm{erf}\left(\frac{t^\prime}{\tau} + \frac{x \nu_x}{c\tau} + i\frac{ \omega_0\tau}{2}\right)\biggr.\\
		& \times \biggl.e^{-2x^2/g^2 -\left(\frac{t^\prime}{\tau} + \frac{x \nu_x}{c\tau}\right)^2 + i \omega_0\tau\left (\frac{t^\prime}{\tau} + \frac{x \nu_x}{c\tau}\right)}\,\mathrm{d}x\biggr]
	\label{eq:integral_we_want}
	\end{split} \\
	I_{2} &= \int\limits_{-\infty}^{\infty} e^{-2x^2/g^2 -2(t^\prime + x \nu_x/c)^2/\tau^2}\,\mathrm{d}x ,
\end{align}
where the identity $\mathrm{erf}^\ast(z) = \mathrm{erf}(z^\ast)$ was used.

The integral $I_1$ may be cast into a generic form by completing the square in the exponent, resulting in
\begin{align}
	I_{1} =\frac{\sqrt{\pi}\tau}{2} Re\left[e^{-\zeta}\int\limits_{-\infty}^{\infty} x\, \mathrm{erf}(a x + b)e^{-(\alpha x + \beta)^2}\,\mathrm{d}x \right]\, ,
	\label{eq:generic_form}
\end{align}
with
\begin{align}
	a &= \nu_x /(c\tau)\\
	b &=  t^\prime/\tau + i\frac{\omega_0\tau}{2}\\
	\alpha &= \sqrt{ \frac{2}{g^2} + \left(\frac{ \nu_x}{c\tau}\right)^2}\\
	\beta &=\frac{a b^\ast}{\alpha} \\
	\zeta &=(b^\ast)^2 + 2\left(\frac{\omega_0\tau}{2}\right)^2 -  \left(\frac{a b^\ast}{\alpha}\right)^2\, .
\end{align}
This integral can be transformed via integration by parts, $\int u\,\mathrm{d}v = uv|_{-\infty}^{\infty}-\int v\,\mathrm{d}u$, using
\begin{align}
	u &= \mathrm{erf}(a x+b) \\
	dv &= x e^{-(\alpha x + \beta)^2}dx \\
	du &= \frac{2a}{\sqrt{\pi}} e^{-(ax+b)^2}dx \\
	v &=(2\alpha^2)^{-1}\left(-e^{-(\alpha x+ \beta)^2} - \beta\sqrt{\pi}\mathrm{erf}(\alpha x + \beta) \right).
\end{align}
Using the large-argument expansion of the error function, the product of the error function terms in $uv$ evaluated at $+ \infty$ is 1, and is 1 at $x=-\infty$, so the difference is zero. The exponential in $v$ evaluated at $\pm\infty$ is zero. Thus the $uv$ term evaluates to zero.

Turning to $-\int v\,\mathrm{d}u$,
\begin{align}
	\begin{split}
	-\int v\,\mathrm{d}u = &\frac{a}{\sqrt{\pi}\alpha^2}%
						\int\limits_{-\infty}^{\infty} e^{-(ax+b)^2} \\
						\times&\left(e^{-(\alpha x + \beta)^2} + \beta\sqrt{\pi}\mathrm{erf}(\alpha x + \beta) \right)\,\mathrm{d}x \, .
	\end{split}
\end{align}
There are two integrals in the previous equation:
\begin{align}
	I_{11} &= \int\limits_{-\infty}^{\infty} e^{-(ax+b)^2 - (\alpha x + \beta)^2}\,\mathrm{d}x \\
	I_{12} &= \beta\sqrt{\pi}\int\limits_{-\infty}^{\infty} \mathrm{erf}(\alpha x+ \beta)e^{-(a x+b)^2}\,\mathrm{d}x\,.
\end{align}
$I_{11}$ straightforward to compute, and is
\begin{align}
	I_{11} = \frac{\sqrt{\pi}}{\sqrt{a^2+\alpha^2}}e^{-(b\alpha - a \beta)^2/(a^2 + \alpha^2)}\, ,
\end{align}
with the restriction that both $a$ and $\alpha$ are greater than zero (which is satisfied here).

$I_{12}$ can be solved using the following tabulated integral, which is Eq.~(13) of section 4.3 in \cite{Ng1969},
\begin{align}
	\int\limits_{-\infty}^{\infty}  \mathrm{erf}(y) e^{-(z y + w)^2}\,\mathrm{d} y = - \frac{\sqrt{\pi}}{z} \mathrm{erf} \left( \frac{w}{\sqrt{z^2 + 1}} \right).
	\label{eq:nbs_formula}
\end{align}
Using Eq.~(\ref{eq:nbs_formula}), in combination with the substitutions,
\begin{align}
	y &= \alpha x + \beta \\
	z &= \frac{a}{\alpha} \\
	w &= b - \frac{\beta a}{\alpha}
\end{align}
the integral can be calculated as
\begin{align}
	I_{12} &=-\frac{\beta \pi}{a} \mathrm{erf}\left( \frac{b \alpha -  a\beta}{\sqrt{\alpha^2 + a^2}}\right)\, .
	\label{eq:nbs_formula_result}
\end{align}
Here, $Re[z^2] = Re[(a/\alpha)^2]$ must be greater than zero, which is satisfied here. Note that $y$ is complex, so the integration limits in Eq.~(\ref{eq:nbs_formula}) should have the imaginary part of $\beta$ added to both. In effect, this changes the integral from over the real line, to a line with a constant imaginary part, parallel to the real line. This change has no effect on the integral for the parameters in this work, as it has been verified using numerical quadrature that Eq.~(\ref{eq:nbs_formula_result}) produces the correct result.

Putting these results together, the integral $I_{1}$ can be calculated as
\begin{align}
	I_{1} = \tau \frac{a}{2 \alpha^2 }%
	\begin{aligned}[t]
	Re&\left\lbrace e^{-\zeta}\left[%
	\frac{\sqrt{\pi}}{\sqrt{a^2 + \alpha^2}} e^{-(b \alpha - a \beta)^2/(a^2 + \alpha^2)}\right. \right. \\
	-&\left.\left.\frac{\pi \beta}{a} \mathrm{erf}\left( \frac{b \alpha -  a\beta}{\sqrt{\alpha^2 + a^2}}\right) \right]%
	\right \rbrace \,.
	\end{aligned}
	\label{eq:I1_semifinal}
\end{align}
Turning to $I_2$, it can be computed as
\begin{align}
	I_2 = \sqrt{\frac{\pi}{2}} \frac{ e^{-\frac{2 (ct^\prime/\nu_x)^2}{g^2 + (c\tau/\nu_x)^2}}}{\sqrt{g^{-2} + (c\tau/\nu_x)^{-2}}}\, .
	\label{eq:I2_final}
\end{align}
Using (\ref{eq:I1_semifinal}) and (\ref{eq:I2_final}) in (\ref{eq:energyflux_I1_I2}), the energy flux can be written as
\begin{widetext}
\begin{align}
	\begin{split}
		E_f^\prime &= \frac{|p_0|^2}{2\rho_0} \left(\frac{\sin\theta_i}{c}  \sqrt{\frac{\pi}{2}} \frac{ e^{-\frac{2 (ct^\prime/\nu_x)^2}{g^2 + (c\tau/\nu_x)^2}}}{\sqrt{g^{-2} + (c\tau/\nu_x)^{-2}}} \right. +\frac{2 \cot\theta_i}{g^2} \frac{\tau \nu_x}{2 c\tau}\frac{1}{2/g^2 + (\nu_x/(c\tau))^2} \mathrm{Re}\left\lbrace\vphantom{\frac{\sqrt{\pi}}{\sqrt{2/g^2 + (\nu_x/(c\tau))^2}}}\right.   \\
		 \times & \exp \left[ \left( \left(\frac{\nu_x}{c\tau\sqrt{2/g^2 + (\nu_x/(c\tau))^2}} \right)^2 -1 \right)\left(\left(\frac{t^\prime}{\tau}\right)^2 - i\omega_0\tau -\left(\frac{\omega_0\tau}{2}\right)^2\right) - 2\left(\frac{\omega_0\tau}{2}\right)^2 \right] \\
	 \times & \left( \frac{\sqrt{\pi}}{\sqrt{2} \sqrt{1/g^2 + (\nu_x/(c\tau))^2}}\exp \left[ -\left( \frac{b\alpha - a\beta}{\sqrt{a^2 + \alpha^2}}  \right)^2  \right] \right. - \left.\left.\left. \frac{\pi(t^\prime/\tau - i \omega_0\tau/2)}{\sqrt{2/g^2 + (\nu_x/(c\tau))^2}} \erf\left[ \frac{b\alpha - a\beta}{\sqrt{a^2 + \alpha^2}} \right]\vphantom{\frac{\sqrt{\pi}}{\sqrt{2/g^2 + (\nu_x/(c\tau))^2}}} \right) \right\rbrace\right)\, , \label{eq:effectiveEnergyFluxFull}
	\end{split}
\end{align}
where the factor
\begin{align}
	 \frac{b\alpha - a\beta}{\sqrt{a^2 + \alpha^2}} &= %
	  \frac{\left(t^\prime/\tau + i \omega_0\tau/2\right)\sqrt{2/g^2 + (\nu_x/(c\tau))^2} - (\nu_x/(c\tau))^2 \frac{t^\prime/\tau - i \omega_0 \tau/2}{\sqrt{2/g^2 + (\nu_x/(c\tau))^2}} }{\sqrt{2}\sqrt{1/g^2 + (\nu_x/(c\tau))^2}}
	  \label{eq:abalphabeta}
\end{align}
\end{widetext}
is not explicitly included in Eq.~(\ref{eq:effectiveEnergyFluxFull}) for space reasons.

It is advantageous to find a simplification for the error function with a complex argument appearing in (\ref{eq:effectiveEnergyFluxFull}). Many software packages only implement the error function with a real argument (e.g. the basic Matlab installation, the GNU C math library, or the GNU C++ cmath library), and an approximation of this special function in terms of elementary functions makes implementing the broadband scattering cross section much simpler numerically. The approximation derived below is very accurate for all parameters investigated here, and has a more succinct representation than Eq.~(\ref{eq:effectiveEnergyFluxFull}).

To see what kind of approximation is appropriate, it is necessary to know if the error function argument is large, or small. To obtain an estimate of the order of magnitude, for this problem at moderate and low grazing angles, $g \gg c\tau/\nu_x$. Therefore, $\alpha \approx a$, and $\alpha^2 + a^2\approx 2 a^2$. The argument of the error function is $q$, and combined with the previous approximation, can be simplified to
\begin{align}
	  q\approx \frac{b - b^\ast}{\sqrt{2}} = \frac{i \omega_0 \tau}{\sqrt{2}}
\end{align}
The largest value of $f_0\tau$ examined in this work is 1.7, resulting in $q\approx 7.5i$. Therefore the magnitude of $q$ is large, and its real part is small compared to its imaginary part. The large-argument asymptotic series approximation is
\begin{align}
	\mathrm{erf}(q) = 1 - \frac{e^{-q^2}}{q\sqrt{\pi}} \left(1 - \frac{1}{2 q^2}+\frac{3}{4 q^4}+ \dots\right).
\end{align}
For $f_0\tau=1.7$, $\exp(-q^2)\approx10^{24}$, so additive factor of 1 outside the parentheses can be safely neglected. The terms inside the parentheses become progressively smaller. The second term is about $10^{-2}$, and the third term is about $10^{-4}$, for $f_0\tau=1.7$, so only retaining the first term is appropriate. If longer pulse lengths are used, this approximation improves, and if shorter pulse lengths are used, then the approximation is poorer. If extremely short pulses are used, then more terms could be kept in the approximation.

Keeping only the largest terms in the power series of $\mathrm{erf}(q)$, the integral $I_1$ is approximately

\begin{align}
	I_{1} 
		  &\approx \sqrt{\frac{\pi}{2}} \frac{ e^{-\frac{2 (ct^\prime/\nu_x)^2}{g^2 + (c\tau/\nu_x)^2}}}{\sqrt{a^2 + \alpha^2}}\frac{\tau a \sqrt{2} }{2\alpha^2} \left(1 + Re\left\{\frac{\beta}{a}\frac{\alpha^2 + a^2}{b \alpha - a \beta} \right\}\right).
\end{align}
where $e^{-\zeta}$ has been combined with the other exponential terms.

This approximation for $I_1$ can be used to obtain a good approximation to the effective energy flux
\begin{align}
	\begin{split}
	E^\prime_f \approx &\frac{|p_0|^2}{2\rho_0 c}L_{eff} e^{-\frac{2 (ct^\prime/\nu_x)^2}{g^2 + (c\tau/\nu_x)^2}}  \sin\theta_i\\
	&\times \left[ 1 +\frac{\cot\theta_i}{\sin\theta_i} \frac{\nu_x }{g^2 \alpha^2}\left(1 + Re\left\{\frac{\beta(\alpha^2 + a^2)}{a(b \alpha - a\beta)}\right\} \right)\right]
	\end{split}
\end{align}
where $L_{eff}$ is defined in the main text, in Eq.~(\ref{eq:Leff}). Substituting in the definitions of $a$, $b$, $\alpha$ and $\beta$, and taking the real part results in Eqs.~(\ref{eq:ef_simplified}) though (\ref{eq:chi}) in the main text.


\end{document}